\documentclass[conference]{IEEEtran}
\IEEEoverridecommandlockouts

\usepackage{cite}
\usepackage{amsmath,amssymb,amsfonts}
\usepackage{algorithmic}          
\usepackage{graphicx}
\usepackage{textcomp}
\usepackage{xcolor}
\usepackage[hyphens]{url}

\usepackage{subcaption}
\usepackage{booktabs}
\usepackage{makecell}
\usepackage{multirow}
\usepackage{threeparttable}
\usepackage{enumitem}
\usepackage[normalem]{ulem}
\usepackage{comment}
\usepackage{stackengine}
\usepackage{tikz}
\usetikzlibrary{arrows.meta,positioning,shapes.geometric,shapes.multipart,fit,calc}

\usepackage{float}

\usepackage[ruled,vlined,linesnumbered]{algorithm2e}
\DontPrintSemicolon
\SetNlSty{\mdseries\upshape}{\mdseries\upshape}{:\,}
\SetAlCapFnt{\small}
\SetAlCapNameFnt{\small}
\SetAlgoNlRelativeSize{-1}

\newcommand{\ADD}[1]{\textcolor{black}{#1}}

\tikzset{
  shape example/.style={
    draw, rounded corners, align=center,
    minimum width=2.4cm, minimum height=1.2cm,
    inner sep=.8pt, font=\small
  }
}

\newcommand{\circnumblue}[1]{%
  \tikz[baseline=(char.base)]{%
    \node[shape=circle, draw=blue, fill=blue, text=white, inner sep=.8pt] (char) {#1};
  }%
}
\newcommand{\circnumorange}[1]{%
  \tikz[baseline=(char.base)]{%
    \node[shape=circle, draw=orange, fill=orange, text=white, inner sep=.8pt] (char) {#1};
  }%
}

\newcommand{\mytitle}{Cerberus}  

\def\BibTeX{{\rm B\kern-.05em{\sc i\kern-.025em b}\kern-.08em
    T\kern-.1667em\lower.7ex\hbox{E}\kern-.125emX}}

\begin{document}

\title{\mytitle{}: Cross-Layer ECC Co-Design\\
for Robust and Efficient Memory Protection}

\author{
\IEEEauthorblockN{
Junhwan Kim\textsuperscript{*\,\P},
Seunghyun Kim\textsuperscript{\textdagger\,\P},
Yesin Ryu\textsuperscript{\ensuremath{\ddagger}},
Saeid Gorgin\textsuperscript{\textsection},
and Jungrae Kim\textsuperscript{*}
}

\IEEEauthorblockA{
\textsuperscript{*}Dept. of Electrical and Computer Engineering,
\textsuperscript{\textdagger}Dept. of Artificial Intelligence,\\
\textsuperscript{\ensuremath{\ddagger}}Dept. of Semiconductor and Display Engineering,
Sungkyunkwan University, Suwon, Republic of Korea\\
\textsuperscript{\textsection}School of Physics, Engineering and Computer Science,
University of Hertfordshire, Hatfield, United Kingdom\\
\{june9918, kshyun0815, seleneyou\}@skku.edu, s.gorgin@herts.ac.uk, dale40@skku.edu\\
}
}
\maketitle


\begin{abstract}

As DRAM scales to higher density and I/O speeds, ensuring data correctness becomes increasingly difficult. Industry has responded with a three-layer stack—on-die ECC (O-ECC), link ECC (L-ECC), and system ECC (S-ECC)—but these layers have evolved independently, often duplicating redundancy, leaving coverage gaps, and occasionally interfering.

We propose \emph{\mytitle{}}, a cross-layer ECC co-design that unifies protection across device, link, and system while preserving each layer’s native role. At its core is an \emph{Encode-Once, Decode-Many (EODM)} architecture: the controller performs a single encoding whose redundancy is reused by L-ECC for immediate write-path detection/retry, by O-ECC for in-device repair on reads, and by S-ECC for strong end-to-end recovery.
\mytitle{} jointly designs complementary parity/syndrome structures, orders decoders and allocates the correction budget to prevent miscorrection amplification, and enables selective correction under tight redundancy constraints. Our evaluations show improved resilience to clustered and peripheral faults while reducing redundancy overhead, underscoring the importance of coordinated cross-layer protection for next-generation memory systems, such as custom HBMs.


\end{abstract}

\begingroup
\renewcommand\thefootnote{}
\footnotetext{\textsuperscript{\P}Both authors contributed equally to this research.}
\endgroup

\section{Introduction}
\label{sec:intro}

As memory technology scales toward ever higher density and bandwidth, ensuring data correctness has become increasingly challenging~\cite{spessot20201t,gong2017scaling,Mutlu2015Main,Hassan2019CROW}. Modern DRAMs integrate billions of tiny capacitors that store femtocoulomb-level charge, operate at sub-nanosecond timing margins, and communicate over multi-gigabit-per-pin I/O links~\cite{ha2018Stanford,park202411,jung2022supply,o2017fine}. These advances enable unprecedented capacity and performance but also amplify vulnerability to faults. A single defective transistor, a marginal sense amplifier, or a transient signaling glitch can corrupt data and compromise large-scale computation~\cite{yim2011hauberk, fiala2012detection}.

The risk is especially pronounced in recent DRAM families such as HBM and LPDDR. These memories map each channel to a single DRAM device (a \emph{single-device-per-channel (SDPC)} organization) to achieve higher bandwidth and lower power.
However, this decision places all data in a single device, effectively putting all eggs in a single basket. Consequently, any device-level defect can compromise the entire channel, whereas traditional multi-device DDR modules can tolerate a failed device through rank-level protection~\cite{Yeleswarapu2020Addressing}.

At the same time, modern data-intensive workloads have reduced the minimum fetch granularity of SDPC memories to 32B (vs. 64B in DDR-based systems) to curb overfetch and improve effective bandwidth. While beneficial for performance, the smaller granularity leaves less room to amortize ECC overheads. As a result, SDPC memories typically allocate only 2\,B of redundancy per 32\,B block (6.25\%), versus 8\,B per 64\,B (12.5\%) in common DDR configurations. This narrow budget leaves little margin for robust protection~\cite{sullivan2023implicit, li2012mage}.

Historically, DRAM reliability relied on a \emph{single layer} of \emph{Error-Correcting Code (ECC)} at the memory controller—\emph{System ECC (S-ECC)}—with simple \emph{Single-Error Correction, Double-Error Detection (SEC-DED)} sufficient when isolated soft errors dominated~\cite{salami2019evaluating}. 
However, as process, voltage, and I/O scaling approach physical limits, fault modes have shifted~\cite{beigi2023systematic, chung2025dram_fault, Jung2023Predicting}. Variability, device aging, and high-speed signaling induce \emph{clustered} and \emph{peripheral} faults spanning 8–32 bits within a single access, alongside transient link corruptions. 

These multi-bit errors exceed SEC-DED’s correction capability and can occasionally evade detection, leading to \emph{Silent Data Corruption (SDC)} that threatens end-to-end correctness.
Beyond the technical implications, these trends also affect manufacturability and product quality. Achieving defect-free DRAM dies has become increasingly difficult as geometries shrink, leading to declining yield. Worse, marginal and intermittent defects often elude production testing, escaping into the field as latent faults\cite{bacchini2014characterization}.

To mitigate both reliability and manufacturability challenges, the DRAM industry has incrementally introduced protection mechanisms at \emph{multiple layers} of the memory hierarchy~\cite{moon2024novel}.
Modern devices now integrate \emph{On-die ECC (O-ECC)} to locally repair manufacturing defects and small-scale faults, thereby improving yield~\cite{patel2019understanding, Patel2020BEER, Kim2023Unity, xie2025breaking}. At the channel boundary, high-speed interfaces employ \emph{Link ECC (L-ECC)} or \emph{Cyclic Redundancy Check (CRC)} to detect transmission errors in real time~\cite{JESD209-DDR5, JESD209-6, JESD270-4}. Meanwhile, memory controllers continue to enhance \emph{S-ECC} to recover from large-granularity failures, such as a dead chip within a module~\cite{yoon2010virtualized}.
Collectively, these mechanisms form a \emph{three-layer ECC hierarchy}—link, on-die, and system—that protects data across distinct physical and temporal domains.

However, these layers have evolved independently, without cross-layer coordination. As a result, they often duplicate redundancy, leave critical fault regions unprotected, and most critically suffer from destructive interference across layers. For example, 
when errors exceed the correction capability of O-ECC, the decoder may \emph{miscorrect} benign bits, increase the number of corrupted bits and turn a pattern S-ECC could have fixed into one it cannot~\cite{criss2020BF}.
L-ECC prevents write-path corruption but operates orthogonally to both O-ECC and S-ECC, limiting end-to-end fault traceability. This fragmented protection model wastes redundancy and struggles to balance reliability, performance, and cost in modern memory systems.

This paper introduces \emph{\mytitle{}}, a novel cross-layer ECC co-design that unifies protection across the link, device, and system layers. \mytitle{} addresses the challenges of current multi-layered memory systems, where O-ECC, L-ECC, and S-ECC are provisioned and managed in isolation. Instead of treating them as separate features, \mytitle{} co-optimizes their roles so that their checks become complementary rather than redundant or destructive, while preserving each layer’s native responsibilities: O-ECC for local repair and yield improvement, L-ECC for real-time detection and retransmission, and S-ECC for robust end-to-end recovery.

At the heart of \mytitle{} is an \emph{Encode-Once, Decode-Many} (EODM) architecture. On writes, the memory controller performs a single encoding step to generate a shared redundancy for all layers. Along the write path, the link layer utilizes the redundancy to detect transmission errors immediately and trigger retransmission when needed (write-side L-ECC). On reads, a bank-group decoder corrects storage-side faults within the device (O-ECC) using that same redundancy. Finally, the controller consumes the full redundancy to deliver strong end-to-end correction and detection against cell, peripheral, and I/O failures (S-ECC and read-side L-ECC). By reusing a single pool of redundancy across layers, EODM improves efficiency and provides seamless coverage without protection gaps observed in HBM configurations.

Realizing this architecture requires novel ECC codes that enable redundancy reuse across layers. The limited redundancy must be allocated across layers to balance their costs and benefits, and reuse must be orchestrated so that any miscorrection at an earlier layer never amplifies the error burden passed to later layers. The scheme must also be sufficiently scalable to support practical co-design and deployment across processor and DRAM vendors.

\mytitle{} achieves these goals by co-designing the generator and parity-check matrices used for ECC. This matrix-level co-design enables interoperability across layers while allowing each layer to fully meet its own protection goals. We also enforce a bounded-fault constraint on O-ECC to ensure that any miscorrection never amplifies the error burden to the system layer. Furthermore, by varying the redundancy within the range feasible for SDPC, \mytitle{} serves as a scalable framework that can be applied across diverse memory systems.

Our evaluation shows that \mytitle{} improves reliability and performance despite lowering HBM’s storage overhead by 33.3\%. With less redundancy than baseline HBM, \mytitle{} still achieves higher overall reliability, and under the same redundancy budget as LPDDR6, it delivers substantially stronger protection across diverse fault locations. In addition, \mytitle{} improves system performance by 0.7\%, mainly by eliminating consecutive encoder stages. Taken together, these results demonstrate that \mytitle{} is a practical and scalable framework for enhancing reliability and performance in future SDPC-enabled systems.

The major contributions of this paper are as follows:
\begin{itemize}
\item We propose the co-design architecture of \mytitle{} based on the EODM structure. By co-designing unified protection across each memory layer, \mytitle{} ensures consistent high reliability across the system while minimizing redundancy.
\item We propose an HBM architecture that applies \mytitle{}. \mytitle{} achieves both system reliability and efficiency by significantly reducing parity storage overhead while maintaining high system-wide reliability against clustered and peripheral faults. 
\item We evaluate \mytitle{} using various workloads, showing that \mytitle{} achieves high performance and lower energy consumption while maintaining superior overall reliability.
\end{itemize}

\section{Background}
\label{sec:background}

This section reviews the fundamentals of ECC and how modern memory systems deploy them across \emph{three layers}. It then summarizes recent observations on DRAM error characteristics that motivate cross-layer ECC designs.

\subsection{Error Correcting Codes}
\label{sec:background:ecc}

Error Correcting Codes (ECC) enable reliable storage and transmission in the presence of physical faults. Most memory systems employ \emph{linear block codes} for their low latency and hardware simplicity~\cite{cojocar2019exploiting}. An $(n,k)$ linear block code maps a $k$-symbol message $\mathbf{m}$ into an $n$-symbol \emph{codeword} $\mathbf{c}$ by appending $r=n-k$ redundancy symbols~\cite{davida1972forward}. A code can be described by a \emph{generator matrix} $G$ with $\mathbf{c}=\mathbf{m}G$ and a \emph{parity-check matrix} $H$ with $H\mathbf{c}^\top=\mathbf{0}$.

Upon receiving $\mathbf{y}=\mathbf{c}+\mathbf{e}$ (where $\mathbf{e}$ is the error vector), the decoder computes the \emph{syndrome} $\mathbf{s}=H\mathbf{y}^\top=H\mathbf{e}^\top$~\cite{holman2001error}. Because $\mathbf{s}$ depends only on the error pattern, the decoder estimates $\hat{\mathbf{e}}$ and reconstructs $\hat{\mathbf{c}}=\mathbf{y}-\hat{\mathbf{e}}$. Decoding outcomes are commonly categorized as: (1) \emph{Detectable and Correctable Error (DCE)}; (2) \emph{Detected but Uncorrectable Error (DUE)} when $\mathbf{s}$ lies beyond the correction radius; (3) \emph{Detected but Miscorrected Error (DME)}, in which the decoder asserts success yet outputs an incorrect codeword; and (4) \emph{Undetectable and Uncorrectable Error (UUE)} with $\mathbf{s}=\mathbf{0}$ despite corruption~\cite{Kim2015Bamboo}. Cases (3) and (4) lead to \emph{Silent Data Corruption (SDC)}, which is the most severe class of memory reliability failure\ADD{~\cite{singh2023silent,dixit2022detecting,agiakatsikas2023impact}.}

The most common main‑memory code is \emph{Single‑Error Correction and Double‑Error Detection (SEC-DED)}~\cite{hamming_code,Hsiao1970SECDED}. An SEC-DED code guarantees correction of any single‑bit error and detection (but not correction) of any double‑bit error. More powerful BCH codes correct $t$ bit errors in an $n$ bit word with redundancy on the order of $t\lceil\log_2(n+1)\rceil$ bits~\cite{bch_code}.

Because DRAM errors frequently occur in bursts or are confined to specific physical regions, systems often employ \emph{non-binary, symbol-based} codes.
\emph{Reed–Solomon (RS)} codes defined over $GF(2^m)$ can correct up to $t$ symbol errors using $2t$ redundant symbols~\cite{rs_code}.
By aligning code symbols with physical fault domains—such as a chip or \emph{Data Pin (DQ)}—RS codes efficiently convert spatially correlated bit errors into a small number of symbol errors, providing strong protection with modest redundancy~\cite{lot_ecc, jeong2020pair, sonawane2019implementation}.

\begin{figure}[t]
    \hfill
    \centering
    \includegraphics[width=\columnwidth]{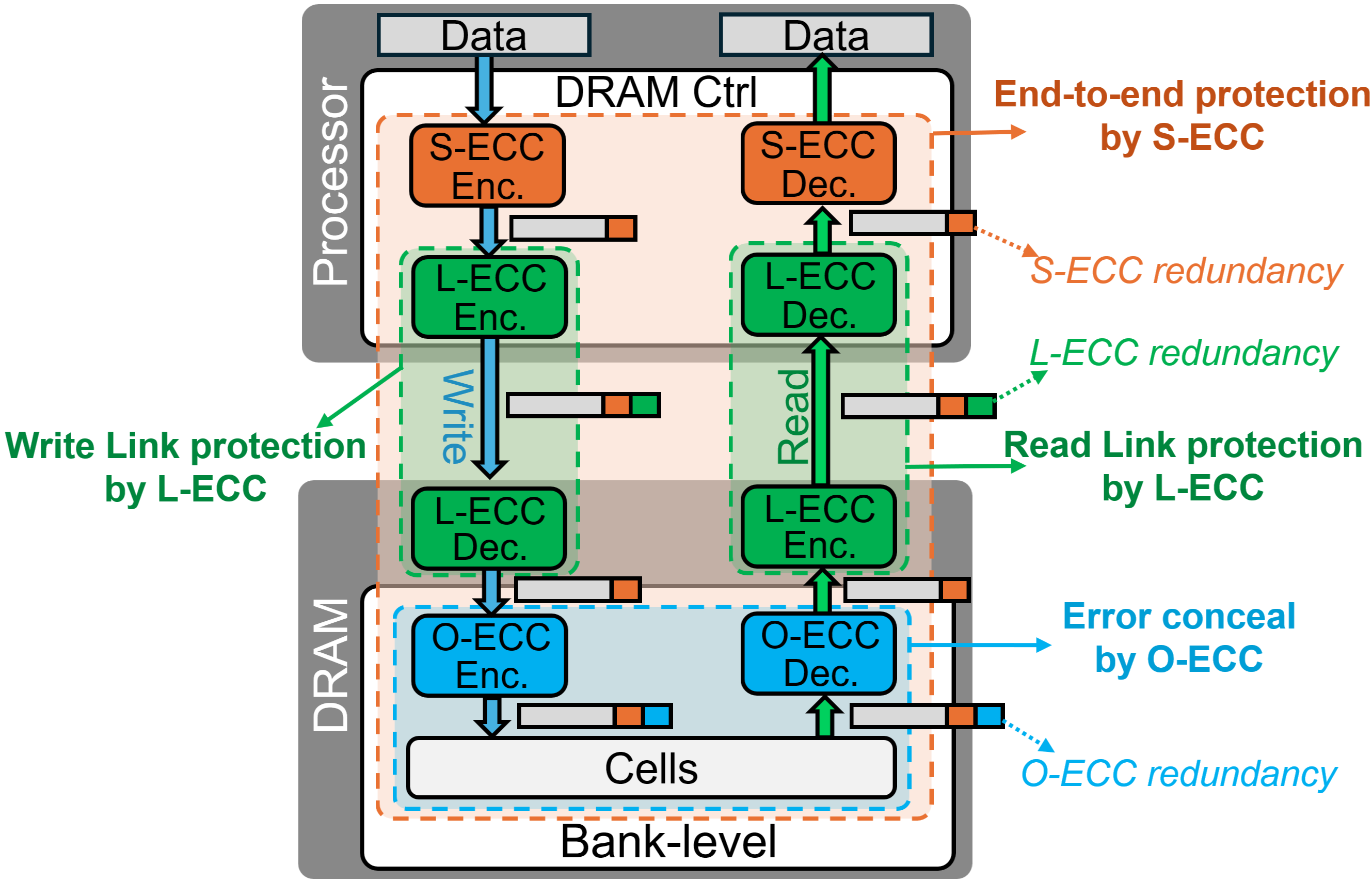}
    \caption{Contemporary DRAM protection based on three ECC layers (S-ECC, O-ECC, and L-ECC)}
    \label{fig:back}
\end{figure}

\subsection{System ECC}
\label{sec:background:secc}

Processors have long employed  ECC to ensure memory and overall system reliability~\cite{chen1984error, hsiao1981reliability}.
The memory controller encodes data on every write and decodes data on every read (orange in Fig.~\ref{fig:back}).
This mechanism—known as \emph{System ECC (S-ECC)} or \emph{rank-level ECC}—detects and corrects errors in both storage and transmission, providing \emph{end-to-end protection} between the processor and memory~\cite{Vadhiraj2020ECC}.

S-ECC implementations vary across DRAM types, generations, and processor vendors.
A standard DDR4 ECC-DIMM provides a $(64+8)$-bit interface, where the additional eight bits store redundancy for each 
64-bit data beat.
Most processors implement SEC-DED, enabling correction of any single-bit error and detection of double-bit errors per beat.
While this configuration is sufficient for random soft errors, large-scale systems require stronger resilience against complete device failures—known as \emph{Single-Device Data Correction (SDDC)} or \emph{chipkill-correct}~\cite{AMDchipkill}.

SDDC implementations vary across vendors and are often proprietary or confidential~\cite{li2022correctable,du2021faultaware}.
Nonetheless, several public designs employ RS-based protection while remaining compatible with standard ECC-DIMMs.
For instance, AMD’s \emph{Chipkill-Correct} constructs 8-bit RS symbols by grouping two 4-bit data beats from each chip and employs two redundant devices for single-chip recovery~\cite{AMDchipkill}.
\emph{Bamboo ECC} forms 8-bit symbols across eight I/O beats, allowing correction of up to four faulty pins ($\approx$ one chip failure)~\cite{Kim2015Bamboo}.
More recently, \emph{Unity ECC} extends this concept to handle both single-symbol and double-bit errors through hybrid decoding~\cite{Kim2023Unity}.

DDR5 reshapes S-ECC design by doubling the burst length (from 8 to 16 beats) and dividing each DIMM into two 32-bit subchannels.
High-reliability ECC-DIMMs allocate eight ECC pins per subchannel, forming $(2\times(32+8))$ interfaces that provide SDDC capability within each subchannel~\cite{JESD209-DDR5}.
In contrast, memory types such as HBM and LPDDR deliver full-channel access through a single device to maximize bandwidth and energy efficiency.
In these \emph{single-device memory channels}, device-level SDDC is infeasible—any device failure compromises the entire channel’s data. Consequently, S-ECC in such systems targets large-granularity burst or link-level faults rather than complete device failures.

\subsection{On-Die ECC}
\label{sec:background:oecc}

Shrinking process technology reduces the charge stored in each cell, increases process variation, and makes transistors more susceptible to wear-out~\cite{Mutlu2015Main}.
To mitigate these errors, modern DRAMs integrate \emph{On-Die ECC (O-ECC)} that repairs faults internally, complementing external S-ECC~\cite{patel2019understanding,Uksong2014Co,Alam2022COMET}.
Each die contains hidden redundancy cells and a compact encoder/decoder that encodes every write and decodes every read within each bank group (blue in Fig.~\ref{fig:back}).
By correcting faults locally, O-ECC effectively converts dies with marginal defects into externally fault-free components, improving manufacturing yield and ensuring component-level reliability beyond warranty thresholds~\cite{Patel2020BEER}.

O-ECC implementations also vary across DRAM types and generations~\cite{chun202016, ryu2023IEEE,oh20143,park2022192}.
In DDR5, where channel data are distributed across multiple chips, devices employ lightweight bit-level SEC schemes with 8 redundant bits per 128 data bits~\cite{JESD209-DDR5}.
In contrast, single-device memory channels such as HBM and LPDDR apply O-ECC at their native 32-byte access granularity.
For instance, HBM4 uses 32 bits of redundancy 
to correct 16-bit–aligned symbol errors, while LPDDR6 applies SEC-DED codes with 16 bits of redundancy
~\cite{JESD270-4,JESD209-6}.

\subsection{Link ECC}
\label{sec:background:lecc}

As I/O speeds reach tens of gigabits per second per pin and signaling voltages continue to scale down, transient transmission errors have become a significant reliability concern~\cite{Kim2016All}.
While S-ECC provides end-to-end protection, it decodes data only on reads and therefore cannot detect \emph{write-path} errors, which can leave corrupted data permanently stored in DRAM. Moreover, S-ECC is often omitted entirely in cost-sensitive or low-power systems to reduce pin count, area, and power consumption~\cite{wang2016use}.

\emph{Link ECC (L-ECC)} protects data during transmission between the memory controller and DRAM (green in Fig.~\ref{fig:back}).
The sender (e.g., the controller on writes) encodes data before transmission, and the receiver verifies it immediately upon arrival, enabling rapid detection—and, in some designs, correction—of transient link errors.
Upon error detection, the receiver can request retransmission, preventing corrupted data from being committed to DRAM.
L-ECC thus serves as the first line of defense in the data path, prioritizing fast detection and low-latency recovery over complex correction.

Different memory types adopt L-ECC in various forms.
DDR5 employs an 8-bit \emph{Cyclic Redundancy Check (CRC)}~\cite{Peterson1961Cyclic} per four DQs, while HBM implements a data-parity bit across every 32 DQs.
LPDDR6 adopts 16-bit parity, configurable for either single-error correction or detection-only operation.

\subsection{DRAM Errors}
\label{sec:dram_errors}

Designing efficient ECC mechanisms requires understanding how DRAM errors manifest in real systems.
While individual DRAM chips are highly reliable, large-scale field studies reveal that aggregate errors exhibit non-negligible rates and distinct patterns~\cite{li2022correctable,wu2024removing,beigi2023systematic, du2021predicting, llama2024hbm3,chung2025dram_fault,mbu_3meza2015revisiting}.
We summarize observations from recent studies as follows:

\subsubsection{Scaling-Induced Cell and Circuit Faults}
As DRAM technology continues to scale down, transient soft errors are increasingly overshadowed by permanent or intermittent faults caused by process variation and device wear-out.
Individual cells have become more susceptible to charge leakage, variable retention time (VRT), and disturbance effects such as row hammering, while peripheral circuits suffer from degraded timing margins and transistor aging~\cite{bacchini2014characterization, spessot20201t, nicolaidis2005design, gong2017scaling}.

\subsubsection{Multi-bit Errors}

Modern DRAMs increasingly exhibit \emph{spatially correlated} multi-bit errors rather than isolated single-bit errors.
Such correlations arise because many peripheral components—such as subwordline (SWL) and subwordline drivers (SWD)—serve multiple adjacent cells~\cite{chung2025dram_fault}.
When one of these shared components fails, it can simultaneously corrupt all of the cells it serves.
Column-related faults typically flip one bit per access, whereas row-related faults can disrupt multiple bits within the same access and thus pose a greater challenge to ECC~\cite{beigi2023systematic}.

The scope of these correlated errors depends heavily on the internal organization of peripheral circuits. Recent characterization of DDR5 devices reveals that most errors remain confined within a small physical region, typically spanning up to 16 bits per access~\cite{chung2025dram_fault}.
In DDR5, each access transfers 8 bits of data from multiple \emph{Memory Array Tiles (MATs)}.
Although MATs are largely independent, adjacent tiles share critical peripheral components—most notably the subwordline driver.
A defect in this shared driver can propagate across MAT boundaries, corrupting both tiles and resulting in up to 16 erroneous bits per access\footnote{Depending on the DRAM architecture, the affected bit width can range from about 8 to 32 bits.}. 
This observation implies that modern ECC mechanisms must be capable of correcting up to 16 clustered errors to maintain high reliability in advanced DRAM technologies.

\subsubsection{Errors Beyond Bank-Groups}
While O-ECC effectively corrects faults within individual bank groups, recent studies reveal that a significant portion of DRAM errors originate beyond these boundaries, such as in device-level peripheral circuits or interconnect paths~\cite{beigi2025ddr5,wu2024removing, mbu_3meza2015revisiting}.
For example, a report on HBM3 devices equipped with integrated O-ECC revealed that, even with O-ECC enabled, a substantial number of error interrupt events were still reported\cite{llama2024hbm3}.
This implies that these errors originated outside the coverage of O-ECC or emerged after the O-ECC stage.
The persistence of such errors indicates that many arise in unprotected regions—e.g., global I/O interfaces, TSV or silicon interposer links—where O-ECC’s correction scope does not apply\cite{Jeon2014Efficient}.
These findings highlight the limitations of O-ECC and reinforce the importance of maintaining end-to-end protection through S-ECC.

\section{Motivation}
\label{sec:motivation}

The previous section outlined three ECC layers, each optimized for a distinct reliability objective:
S-ECC provides end-to-end protection and strong overall reliability,
O-ECC conceals errors and improves manufacturability~\cite{nair2016XED, Patel2020BEER, patel2019understanding, Alam2022COMET}, and
L-ECC enables early detection of link errors.
This section examines how commercial DRAMs combine these layers in practice.
Although each layer is effective in isolation, their ad-hoc integration frequently results in redundant coverage, inefficient use of redundancy, and—paradoxically—reduced overall reliability, motivating the need for a cross-layer ECC framework.

\subsection{ECCs in DDR-based Systems}
\label{sec:motivation:ddr}
Although \mytitle{} targets single-device memory architectures such as HBM and LPDDR, it is instructive to first examine DDR-based systems.
High-reliability platforms, including supercomputers, have developed sophisticated reliability mechanisms for DDR memory, and the distribution of data across multiple DRAM devices within a DIMM inherently enables strong SDDC protection~\cite{AMDchipkill}.

A typical DDR5 system employs three ECC layers configured as follows:
(1) \emph{S-ECC}, implemented with 25\% additional devices to provide SDDC-level protection;
(2) \emph{O-ECC}, adding 6.25\% cell-area overhead (8 parity bits per 128-bit data word) for SEC within each device; and
(3) \emph{L-ECC}, introducing a 12.5\% transfer overhead (two additional beats per 16-beat burst) to provide CRC16-based link error detection.
Together, these layers incur approximately 32.8\% storage overhead (from S-ECC and O-ECC) and 40.6\% transfer overhead (from S-ECC and L-ECC), highlighting the inefficiency of independently managed ECC layers~\cite{JESD209-DDR5}.

\begin{figure}[t]
\centering
\subfloat[Miscorrection by SEC O-ECC without BF]{
\includegraphics[width=0.7\columnwidth]{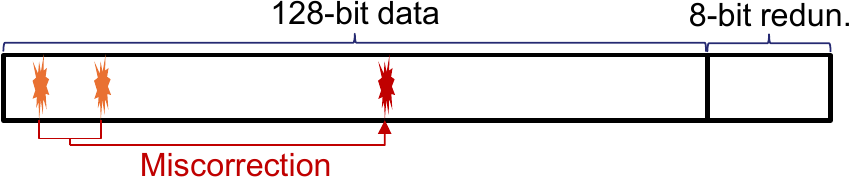}
\label{fig:boundedfault1}
}

\subfloat[Miscorrection by SEC O-ECC with BF]{
\includegraphics[width=0.7\columnwidth]{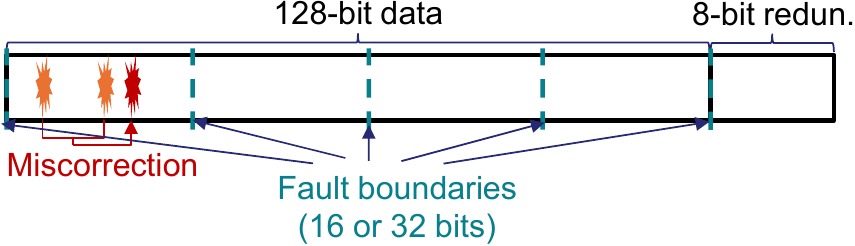}
\label{fig:boundedfault2}
}

\subfloat[Parity-check matrix enforcing BF behavior]{
\includegraphics[width=0.8\columnwidth]{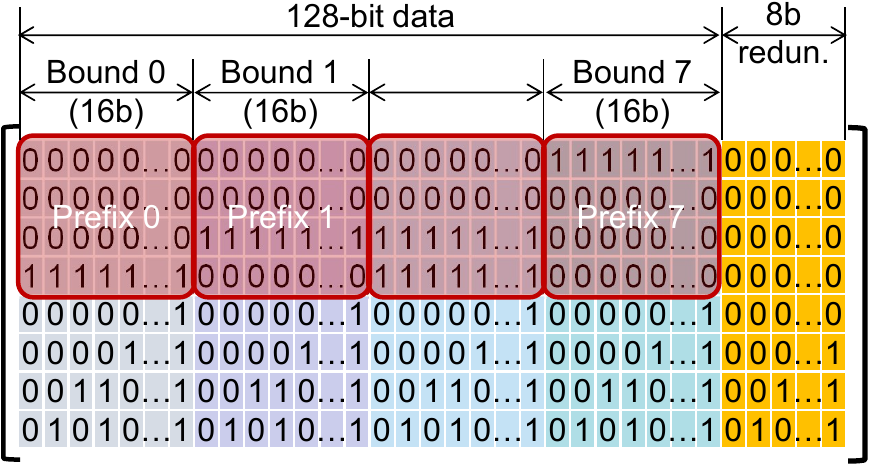}
\label{fig:boundedfault3}
}
\caption{Bounded-fault design for SEC O-ECC in DDR5}
\label{fig:boundedfault}
\vspace{-15pt}
\end{figure}

Despite these costs, reliability can degrade due to \emph{miscorrections}~\cite{Alam2022COMET,Halbert2017Memory}. 
When two bits fail, an O-ECC configured as SEC may wrongly flip a third bit (Fig.~\ref{fig:boundedfault1}).
If this new error falls in a different S-ECC symbol, the number of erroneous symbols may exceed S-ECC’s correction capability, producing an uncorrectable fault.
\ADD{
Such miscorrections can occur at nontrivial rates. Under an SEC O-ECC + SEC-DED S-ECC stack, prior work reports that O-ECC miscorrects $\approx 45\%$  of double-bit errors (DBEs) into triple-bit errors, and that S-ECC then miscorrects these triple-bit errors as single-bit errors in 
$\approx 55\%$ of the cases, causing SDC~\cite{Alam2022COMET}. It also estimates that SDC can occur once per 3 million accesses when the DRAM raw error rate is $10^{-4}$.
}


To prevent such cross-layer interference, DDR5 enforces the \emph{Bounded Fault (BF)} rule, which restricts each correction to a small spatial region (Fig.~\ref{fig:boundedfault2})—typically 16 bits from an I/O pin.
O-ECC must ensure that miscorrections remain within the boundary region~\cite{criss2020BF}.
This behavior is guaranteed by designing the parity-check matrix $H$ such that no sum of columns within a region equals any column outside that region.
In one such parity-check matrix (Fig.~\ref{fig:boundedfault3}), columns within one region share a prefix: odd-bit errors preserve it (staying local), while even-bit sums cancel to zero, mapping to non-data space.

The BF rule effectively isolates O-ECC from S-ECC, allowing intra-device correction without propagating faults.
However, each layer still maintains separate redundancy, inflating total storage overhead, and the BF layout constrains S-ECC’s symbol organization to Bamboo-ECC-like groupings\cite{Kim2015Bamboo}.

\subsection{ECCs in HBM-based Systems}

HBM transfers data through a single device, allowing the granularity of O-ECC to align directly with that of S-ECC.
In HBM4, each pseudo-channel protects 32 bytes of data with 2 bytes of S-ECC, 4 bytes of O-ECC, and 1 byte of L-ECC redundancy~\cite{JESD270-4}.
This configuration emphasizes on-die correction by allocating more redundancy bits to O-ECC, allowing it to correct up to 16 faulty bits per block.
Such strong on-die protection effectively suppresses scaling-induced faults, especially those originating from peripheral circuits such as subwordline drivers.
Meanwhile, HBM4 maintains high bandwidth by transmitting L-ECC through a dedicated sideband pin, leaving the main data interface fully utilized.
To balance total redundancy, HBM4 limits S-ECC to 2 bytes per 32-byte data block.
This limited budget can be used for either an 
ECC (e.g., SEC-DED) or an \emph{Error Detecting Code (EDC)} (e.g., CRC16).
In practice, most systems adopt CRC since its misdetection probability ($\approx$0.002\%) is nearly two orders of magnitude lower than that of SEC-DED ($\approx$0.4\%), substantially reducing the risk of SDC~\cite{ryu2023IEEE}.

Despite these refinements, HBM’s reliability remains constrained by several structural limitations.
First, because O-ECC performs symbol-based correction without a bounded-fault constraint, a miscorrection can produce errors beyond the correction capability of S-ECC,
leading to an unrecoverable condition at the system level.
Second, O-ECC can detect most uncorrectable events with high probability ($\approx$99.97\%),
but the results are conveyed to the system level only through a limited severity (SEV) pin~\cite{JESD238, JESD270-4}.
As a result, the system lacks sufficient cross-layer visibility,
and S-ECC must rely solely on the limited error information interpreted by O-ECC (e.g., CE or UE signals),
performing decoding under incomplete awareness.
Consequently, the system becomes exposed to errors undetected by S-ECC,
increasing the risk of missing multi-symbol errors.
Third, O-ECC’s protection scope is confined to the internal array and peripheral circuitry; link and I/O interface errors remain outside its coverage.
This issue is particularly severe in HBM, where Through-Silicon Vias (TSVs) introduce new fault modes. Because a TSV fault lies outside O-ECC’s protection scope, detection-based S-ECC alone cannot adequately handle this error.
Recent field studies of HBM3 devices~\cite{llama2024hbm3} corroborate these issues, showing that a substantial number of errors are still detected by S-ECC, indicating that many faults occur beyond the reach of O-ECC.

\begin{table*}[h]
\caption{Comparison of ECC schemes}
\centering
\label{tab:comparisonECC}
\resizebox{\textwidth}{!}{
\begin{tabular}{l|l|c|c|c|c|c|c|c}
    \Xhline{3\arrayrulewidth}
    
    \multicolumn{2}{c|}{} 
    & \multicolumn{2}{c|}{Single-layer}
    & \multicolumn{4}{c|}{Multi-layer}
    & Cross-layer \\
    
    \cline{3-9}
    
    \multicolumn{2}{c|}{} 
    & DUO
    & Unity ECC
    & \makecell{\ADD{LPDDR6/}\\\ADD{SEC-DED}}
    & \makecell{\ADD{LPDDR6/}\\\ADD{CRC}}
    & \makecell{\ADD{HBM4/}\\\ADD{SEC-DED}}
    & \makecell{\ADD{HBM4/}\\\ADD{CRC}}
    & \mytitle{} \\

    \Xhline{3\arrayrulewidth}

    \multirow{3}{*}{Total}
    & Data bits
    & \multicolumn{2}{c|}{512}
    & \multicolumn{5}{c}{256} \\

    \cline{2-9}
    
    & Storage overhead
    & 19.5\% (100b)
    & 25\% (128b)
    & \multicolumn{2}{c|}{12.5\% (32b)}  
    & \multicolumn{2}{c|}{18.8\% (48b)}  
    & 12.5\% (32b) \\

    \cline{2-9}
    
    & Transfer overhead
    & 19.5\% (100b)
    & 25\% (128b)
    & \multicolumn{2}{c|}{12.5\% (32b)}  
    & \multicolumn{2}{c|}{15.6\% (40b)}  
    & 12.5\% (32b) \\

    \Xhline{3\arrayrulewidth}
    
    \multirow{2}{*}{S-ECC} 
    & Bit config.
    & 512 + 100
    & 512 + 128
    & \multicolumn{4}{c|}{256 + 16}
    & 256 + 16 + 16 \\
    
    \cline{2-9}
    & ECC
    & RS(76,64)
    & SSC+DEC
    & SEC-DED
    & CRC
    & SEC-DED
    & CRC
    & SSC+DEC \\

    \hline

    \multirow{2}{*}{O-ECC} & Bit config.
    & \multicolumn{2}{c|}{{\multirow{2}{*}{N/A}}}
    & \multicolumn{2}{c|}{272 + 16}  
    & \multicolumn{2}{c|}{272 + 32}  
    & 272 + 16\\
    
    \cline{2-2} \cline{5-9}
    
    & ECC
    & \multicolumn{2}{c|}{}
    & \multicolumn{2}{c|}{SEC-DED}   
    & \multicolumn{2}{c|}{16b SSC}   
    & SEC-DED\\

    \hline

    \multirow{2}{*}{L-ECC} & Bit config.
    & \multicolumn{2}{c|}{{\multirow{2}{*}{N/A}}}
    & \multicolumn{2}{c|}{272 + 16}  
    & \multicolumn{2}{c|}{272 + 8}   
    & 272 + 16\\

    \cline{2-2} \cline{5-9}

    & ECC
    & \multicolumn{2}{c|}{}
    & \multicolumn{2}{c|}{SEC or EDC}  
    & \multicolumn{2}{c|}{Parity}      
    & EDC \\
    
    \Xhline{3\arrayrulewidth}
    
    \multicolumn{2}{c|}{Early Detection}
    & \multicolumn{2}{c|}{No}
    & \multicolumn{5}{c}{Yes} \\

    \hline
    
    \multicolumn{2}{c|}{Error Concealment}
    & \multicolumn{2}{c|}{No}
    & \multicolumn{5}{c}{Yes} \\
    
    \hline

    \multicolumn{2}{c|}{Bounded Fault}
    & \multicolumn{6}{c|}{No}
    & \multicolumn{1}{c}{Yes} \\
    
    \hline
    
    \multicolumn{2}{c|}{Correction}
    & \multicolumn{2}{c|}{High}
    & \multicolumn{2}{c|}{Low}
    & Medium   
    & Low      
    & High \\

    \hline

    \multicolumn{2}{c|}{Detection}
    & Very High
    & High
    & Low      
    & High     
    & Medium   
    & High     
    & High \\    
    
    \Xhline{3\arrayrulewidth}
\end{tabular}}
\vspace{-10pt}
\end{table*}

\subsection{ECCs in LPDDR-based Systems}

Similar to HBM, LPDDR employs a single-device memory channel architecture, allowing S-ECC and O-ECC to operate at the same granularity.
In LPDDR6, each pseudo-channel protects a 32-byte data block using 2 bytes of S-ECC, 2 bytes of O-ECC, and 2 bytes of L-ECC redundancy. The O-ECC employs its 2-byte budget to implement SEC-DED for the 32-byte data block and its associated S-ECC redundancy.
Externally, a subchannel transfers 36 bytes per burst across 12 I/O pins over 24 beats—32 bytes of user data, 2 bytes of S-ECC, and 2 bytes carrying either L-ECC redundancy or \emph{Data Bus Inversion (DBI)} information.
Overall, this configuration introduces about 12.5\% storage and transfer overhead, a cost considered acceptable in mobile devices~\cite{JESD209-6}.

Despite these protections, LPDDR6 still provides only moderate reliability due to two key limitations: \emph{miscorrection propagation} and \emph{limited correction capability}. Because LPDDR6 lacks Bounded Fault (BF) protection, an O-ECC miscorrection can turn a triple-bit error into a quadruple-bit corruption, increasing the burden on S-ECC. In addition, SEC-DED-based O-ECC can correct only single-bit errors, so it faces a fundamental limitation when handling multi-bit errors that often arise in peripheral circuitry. The same constraint applies to S-ECC when it uses the same redundancy budget as O-ECC. If S-ECC is configured as SEC-DED, it also corrects only single-bit errors and cannot handle multi-bit errors. Alternatively, configuring S-ECC as an 8-bit Single Symbol Correction (SSC) allows correction of up to 8-bit clustered errors, but its detection capability is limited ($\approx$86.7\%).

\section{Prior Work}
\label{sec:prior}


The inefficiency and limited reliability of multi-layer memory protection in commercial DRAMs have inspired numerous academic efforts to redesign ECC architectures.
\ADD{Prior work largely follows two directions: (i) schemes that consolidate protection into a strengthened S-ECC, and (ii) schemes that explicitly coordinate layers and leverage cross-layer information.}
This section summarizes representative research along these lines.

\begin{figure}[t]
    \centering
    \includegraphics[width=\columnwidth]{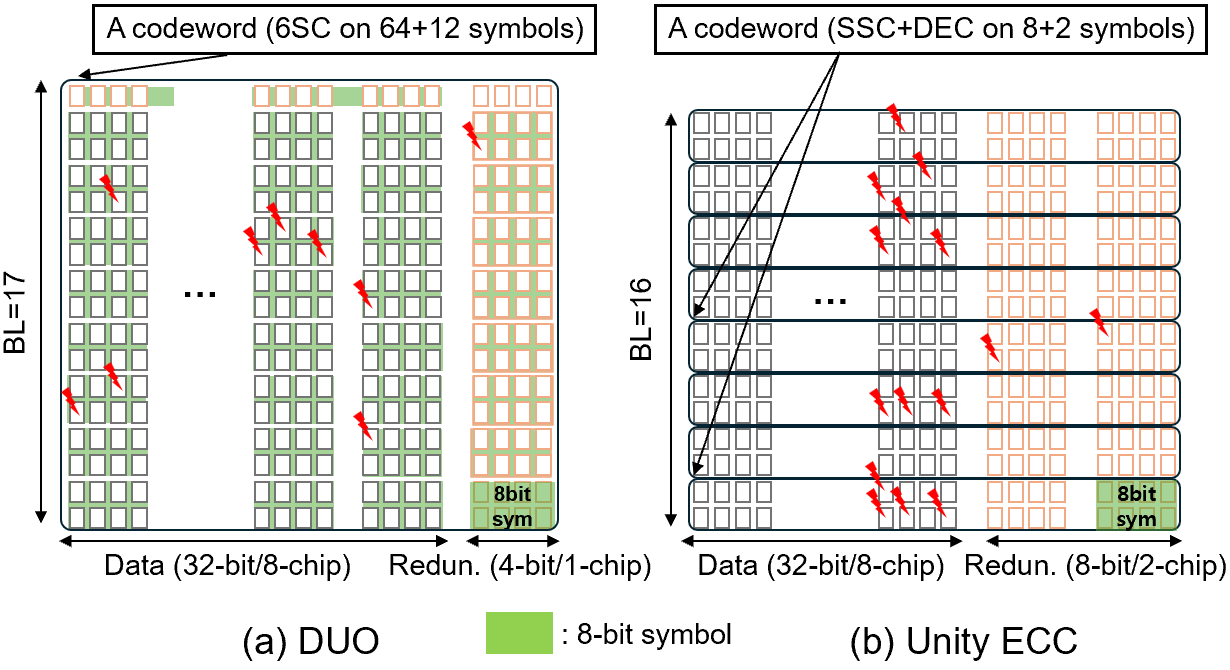}
    \caption{The single-layer ECC configurations of DDR5}
    \label{fig:Prior}
    \vspace{-5pt}
\end{figure}

\subsection{\ADD{Stronger System ECC}}

\subsubsection{DUO} 

DUO~\cite{gong2018DUO} bypasses on-die ECC and repurposes its internal redundancy at the system level.
By forwarding redundant bits to the host through additional transfer beats, DUO extends S-ECC into a longer symbol-based codeword (e.g., RS(76,64)), providing SDDC protection.
A portion of this redundancy is reserved as \emph{on-chip redundancy parity}, which supports burst-erasure correction and verification, enabling recovery even when a complete chip failure coincides with multiple hard defects (Fig.~\ref{fig:Prior}a).

\subsubsection{Unity ECC} 
Unity ECC~\cite{Kim2023Unity} presents a single-layer S-ECC framework capable of correcting both double-bit errors and single-chip failures. 
By allowing S-ECC to handle both frequent bit faults and rare device-level errors, Unity ECC eliminates the need for on-die correction.
This simplification reduces latency, power, and area overheads associated with O-ECC while maintaining robust end-to-end protection (Fig.~\ref{fig:Prior}b).

\subsubsection{\ADD{Dual-Axis ECC}}
\ADD{Dual-Axis ECC~\cite{Jung2024DA} protects both storage and transfer errors via two-orientation decoding: vertical symbols correct storage faults, while horizontal overlays correct common DQS-induced transfer errors by reusing spare syndromes, eliminating L-ECC  overhead and reducing retransmissions without extra redundancy.}

\subsection{\ADD{Cross-layer Collaboration}}

\subsubsection{\ADD{XED}}
\ADD{XED~\cite{nair2016XED} proposes cooperative interaction between S-ECC and O-ECC to achieve SDDC protection. It exposes on-die ECC detection outcomes to the system through a predefined \emph{catch-word} interface, enabling S-ECC to perform erasure-based decoding with improved correction efficiency in configurations with multiple DRAM devices per channel.}

\subsubsection{\ADD{HARP}}
\ADD{HARP~\cite{patel2021harp} improves reliability through profiling-based repair under O-ECC. It leverages O-ECC’s information (e.g., via decode-bypass and/or correction-event reporting) to profile vulnerable locations and then applies repair actions to mitigate future errors at those locations. Its benefit depends on profiling coverage and available repair resources.}

Table~\ref{tab:comparisonECC} summarizes industry practice and two representative academic designs (DUO and Unity ECC) \ADD{for single-device-per-channel organizations}.
Collectively, these academic approaches demonstrate the potential of single-layer ECC to improve system reliability and reduce redundancy. However, they overlook a critical industry requirement—\emph{error concealment within DRAM devices}. To preserve product quality perception, DRAM vendors deliberately mask internal fault behavior, reporting only a limited set of vulnerable regions rather than
exposing raw error counts~\cite{JESD209-5C}. Although academic proposals
enhance transparency and end-to-end reliability, their exposure
of device-level errors to the host conflicts with this industrial
practice, raising concerns over warranty obligations and vendor accountability.

\section{\mytitle{}}
\label{sec:main}

This section introduces \mytitle{}, a unified, \emph{cross-layer} ECC framework that delivers high reliability at low overhead with robust end-to-end protection. Unlike conventional multi-layer schemes in which each layer is designed and operated independently, \mytitle{} \emph{co-designs} the generator and parity-check matrices across layers to ensure interoperability. This interoperability allows a single encoder to produce redundancy shared across layers—each serving a distinct purpose (link protection, error concealment, or ultra-high reliability)—while preventing destructive cross-layer interference (e.g., avoiding miscorrections that amplify errors for the next layer). The following sections present the \mytitle{} architecture and its operational flow.

\subsection{Architecture}
\label{sec:main:architecture}

\mytitle{} targets single-device memory channels in which the channel transfer unit matches the device’s internal access unit (e.g., HBM, LPDDR). In this paper, we focus on an HBM configuration with a 256-bit (32B) access unit and a total redundancy budget of $12.5\%$ (4B per 32B data block). The 32B access unit aligns with current HBM practice, whereas the redundancy budget is lower than the HBM baseline (18.8\%). Despite using less redundancy, \mytitle{} delivers stronger protection, making it suitable for custom HBMs and future HBM generations.

\begin{figure}[t]
    \hfill
    \centering
    \includegraphics[width=0.95\columnwidth]{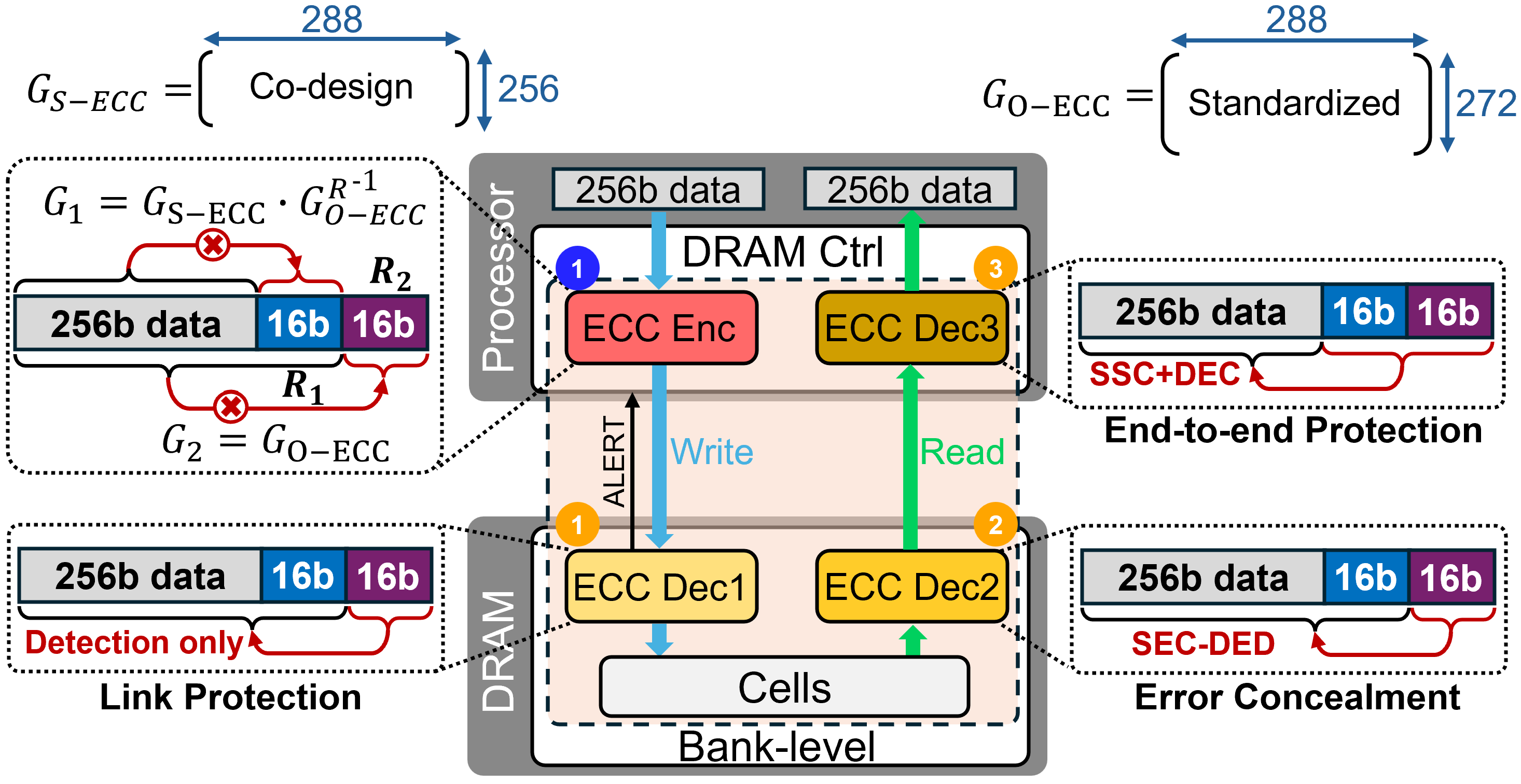}
    \caption{An overview of \mytitle{}}
    \label{fig:main_overview}
    \vspace{-10pt}
\end{figure}

Fig.~\ref{fig:main_overview} illustrates the overall architecture. The framework consists of a single shared encoder (\circnumblue{1}) and three layer-specific decoders (\circnumorange{1}, \circnumorange{2}, \circnumorange{3}) that cooperate along the data path. The 32-bit redundancy generated once by the encoder is reused—wholly or partially—by the following decoding layers:
(1) the \emph{Link Layer} (LL), which provides early detection of write-path link errors;
(2) the \emph{Device Layer} (DL), which performs on-die error correction and concealment; and
(3) the \emph{System Layer} (SL), which ensures strong end-to-end, system-level reliability.
Each layer interprets the shared 32-bit redundancy according to its role. The LL utilizes 16 bits exclusively for error detection; the DL reuses that same 16-bit portion for bit-level single-error correction (SEC) within the die; and the SL leverages the full 32 bits to perform symbol-based correction and detection. 
This 
\emph{Encode-Once, Decode-Many} (EODM) organization with hierarchical redundancy reuse eliminates repeated encoding stages, reduces latency, storage overheads, and preserves seamless protection coverage across all layers without reliability gaps.

\subsection{\mytitle{} Operations}
\label{sec:operations}

\subsubsection{Write Operation}

On a write, the encoder (\circnumblue{1}) takes 256-bit user data $D$ and appends 32 bits of redundancy composed of $R_1$ and $R_2$ (16 bits each).
This redundancy is produced by a generator matrix $G_{\mathrm{S\text{-}ECC}}$, which is a product of two sub-matrices, $G_1$ and $G_2$ ($G_{\mathrm{S\text{-}ECC}} = G_1 \cdot G_2$).
$G_1$ first maps 256-bit $D$ to a 272-bit intermediate codeword ($D$+$R_1$); $G_2$ then maps this 272-bit word to a 288-bit final codeword (($D$+$R_1$)+$R_2$). Although the description separates these as two encoding stages for conceptual clarity, practical implementations can perform both mappings in a single step using the composite matrix $G_{\mathrm{S\text{-}ECC}}$.

The host omits conventional write-path L-ECC and transmits the 288-bit codeword (($D$+$R_1$)+$R_2$) directly to DRAM. On the DRAM side, the first decoder (\circnumorange{1}) replaces L-ECC by verifying the link integrity of (($D$+$R_1$)+$R_2$) using $R_2$ via the parity-check matrix $H_2$ (the dual of $G_2$). Upon a mismatch, the decoder does not attempt correction but instead requests retransmission (e.g., via a conventional ALERT signal). The 16-bit redundancy provides early detection of transfer errors with a detection capability comparable to CRC16. 
If verification succeeds, the validated codeword (($D$+$R_1$)+$R_2$) is stored in DRAM with its redundancy preserved for subsequent device- and system-level reuse.

\subsubsection{Read Operation}

On a read, the second decoder (\circnumorange{2}) operates inside the bank group to perform \emph{on-die} error correction and concealment.
The device validates (($D$+$R_1$)+$R_2$)
using the same parity-check matrix $H_2$, but this time, it applies the correction (SEC) when needed.
This ensures that single-bit errors are corrected within the device, preventing their propagation to the memory controller.

Importantly, $H_2$ is designed to satisfy a \emph{bounded-fault} constraint at the 16-bit symbol granularity: any miscorrection that arises from multiple flipped bits is \emph{confined} to the originally faulty 16-bit symbol and cannot corrupt additional symbols. Consequently, device-level actions never increase symbol-level error severity.
After local correction, the DRAM bypasses read-path L-ECC generation and instead forwards the corrected 288-bit codeword (($D$+$R_1$)+$R_2$) to the controller.

At the controller, the third decoder (\circnumorange{3}) provides system-level protection using both redundancy fields \((R_1,R_2)\). Leveraging the full 32-bit redundancy, it delivers SSC+DEC capability: it corrects either a single 16-bit symbol error (e.g., from a device-level miscorrection or critical peripheral-circuit faults) or two bit errors located in distinct symbols (e.g., overlapping storage/transfer errors). For more severe patterns, it still provides safe detection with probability \(99.97\%\). It also extends correction beyond bank-group boundaries and replaces read-side L-ECC.
\ADD{When the decoder detects an uncorrectable error, it issues a single retry. If the retry is still uncorrectable, it reports a DUE. Otherwise, it treats the first event as a transient read-link or peripheral error and forwards the retry result, as in conventional L-ECC retry.}

Collectively, \mytitle{} delivers SSC+DEC correction to the system, SEC correction to DRAM devices, and CRC16-level detection for link protection, meeting the distinct requirements of DRAM reliability and manufacturability.
Compared to typical HBM designs, \mytitle{} offers stronger end-to-end correction at \(12.5\%\) overhead (vs.\ \(18.8\%\)) and is not confined to a single bank group. Compared to LPDDR-style protection, it matches the \(12.5\%\) overhead but upgrades from bit-level SEC to symbol-aware SSC+DEC at the host, while retaining on-die SEC and link detection for full end-to-end coverage.


\subsection{Cross-Layer ECC Collaboration}
\label{sec:main:ecc}

This architecture raises a key question: how can the system and device layers collaborate so that redundancy provisioned for the system layer is reusable in the device layer? We address this question in the following subsections by describing our device/link-layer, system-layer, and cross-layer ECC designs.


\begin{figure*}[t]
\centering
\subfloat[Parity-check matrix of device/link layer]{
\includegraphics[width=0.8\textwidth]{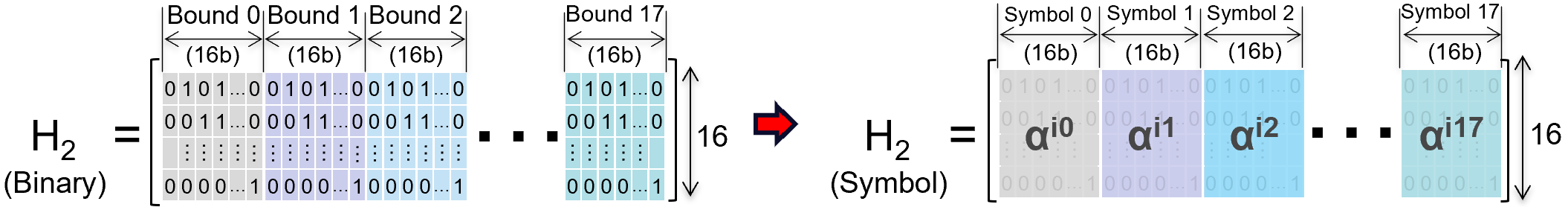}
\label{fig:S-ECC1}
}

\subfloat[Parity-check matrix of system layer]{
\includegraphics[width=\textwidth]{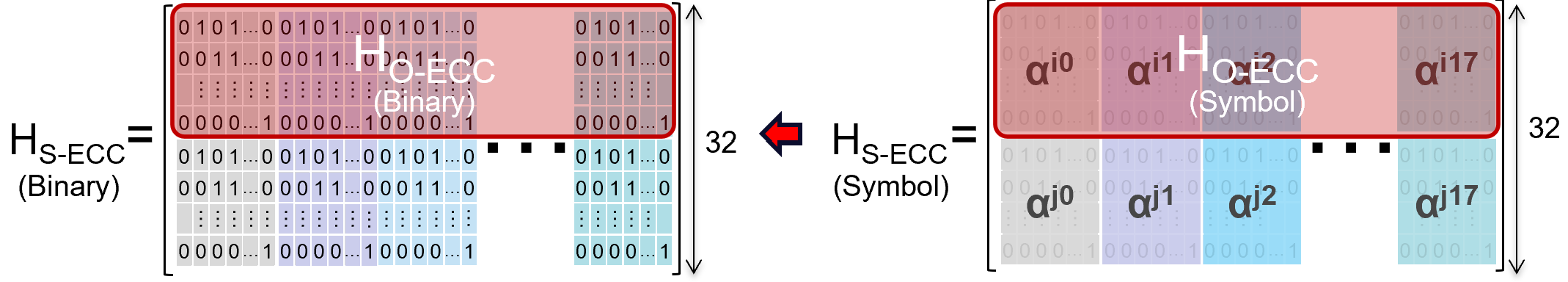}
\label{fig:S-ECC2}
}
\caption{The parity-check matrices of \mytitle{} for cross-layer design}
\label{fig:S-ECC}
\vspace{-10pt}
\end{figure*}

\subsubsection{Device/Link Layer}
We begin by describing the decoding mechanism at the device and link layers.
These layers receive 288-bit data (($D$+$R_1$)+$R_2$) and use $R_2$ to detect and correct errors in $D$ and $R_1$, resulting in a parity-check matrix $H_2$ with dimensions $16\times288$.

The device layer targets SEC-DED with a bounded-fault constraint at 16-bit granularity. We choose SEC based on observations from HBM3/4, where strong on-die correction can trigger severe miscorrections and still leave the system exposed to faults outside the bank group. Such miscorrections then force the system layer to provision additional correction strength and redundancy. For example, an O-ECC using 8-bit SSC with 2-symbol redundancy can miscorrect a two-symbol error into a three-symbol corruption, which would require 6-symbol redundancy in S-ECC for full recovery. Moreover, because error sources are not confined to a single bank group, strong intra-bank protection is helpful but offers limited reliability beyond the bank group. Therefore, we design the device layer to handle frequent small-scale bit errors, while the system layer is responsible for rarer but more severe faults.



To achieve SEC-DED with bounded-fault recovery, $H_2$ must satisfy the following conditions:
\begin{itemize}
\item{Every column must be non-zero.}
\item{SEC: All columns must be unique.}
\item{DED: The sum of any two distinct columns must not be equal to any other column.}
\item{Bounded fault: The sum of columns within any 16-column region must not match any column belonging to another region.}
\end{itemize}

Additionally, since $H_2$ is also used in the link layer, it must be able to detect frequent transfer errors. To achieve this, we design $H_2$ to meet both the device-layer constraints and the CRC8 requirements. Unlike CRC16, CRC8 does not guarantee detection of all burst errors of length 9–16; however, the combined 16-bit redundancy still provides a random-error detection probability of $1-2^{-16}$ ($\approx$99.998\%). Detecting 16-bit burst errors is especially important in DDR systems, where the burst length is typically 16, whereas in memory systems with shorter bursts (e.g., HBM4 with burst-8), this level of detection is less critical. To realize CRC8, $H_2$ must satisfy the condition that any eight consecutive columns are linearly independent.

As a result, we use $H_2$ to protect both the device and link layers, implementing SEC-DED with a bounded-fault property while also meeting CRC8 and providing CRC16-level detection capability.

\subsubsection{System Layer}

The system layer is responsible for ensuring end-to-end reliability by detecting and correcting errors that may have propagated through the device and link layers. This layer is designed to provide robust protection against severe, rare errors that escape the correction capabilities of the device and link layers. The system layer achieves this by leveraging the redundancy provided by the device layer (via $R_1$ and $R_2$) to perform symbol-based error correction, leading to a parity-check matrix ($H_{\mathrm{S\text{-}ECC}}$) with dimensions of $32\times288$.

To achieve this, the system layer is designed to correct single 16-bit symbol errors or double-bit errors (SSC+DEC). This dual capability addresses both clustered errors caused by malfunctioning peripheral circuits (e.g., subwordline driver failures) and frequent, random bit errors that may occur simultaneously. The goal of the system layer is to maintain high levels of protection with minimal additional redundancy overhead, thereby ensuring end-to-end reliability without significant performance or storage costs.

To achieve SSC+DEC, $H_{\mathrm{S\text{-}ECC}}$ must satisfy the following conditions:
\begin{itemize}
\item{Every column must be non-zero.}
\item{SSC: The sums of all symbol-aligned columns are unique.}
\item{DEC: The sums of any two columns are unique.}
\item{SSC+DEC: All sums from properties 2 and 3 should
be unique (apart from double-bit errors in the same symbol,
which are considered symbol errors).}
\end{itemize}

For effective collaboration between the device/link layers and the system layer, $H_{\mathrm{S\text{-}ECC}}$ must satisfy a single condition, which we discuss in the next section.



\subsubsection{Cross Layer}
This section describes how $H_{\mathrm{S\text{-}ECC}}$ can be decomposed into $H_1$ and $H_2$ while encoding only once, and presents the single condition under which any 256-bit granularity DRAM can adopt the \mytitle{} framework. We begin by describing the shared-encoder generator matrix, $G_{\mathrm{S\text{-}ECC}}$.


We design $G_{\mathrm{S\text{-}ECC}}$ to satisfy $G_{\mathrm{S\text{-}ECC}} = G_1 \cdot G_2$. For such a $G_1$ to exist, each row of $G_{\mathrm{S\text{-}ECC}}$ must be expressible as a linear combination of the rows of $G_2$. In other words, the row space of $G_{\mathrm{S\text{-}ECC}}$ must be contained within the row space of $G_2$ (i.e., \( \operatorname{row}(G_{\mathrm{S\text{-}ECC}}) \subseteq \operatorname{row}(G_2) \)).
To ensure this condition in the parity-check matrix domain, we utilize the relationship between the generator matrix $G$ and the parity-check matrix $H$ (e.g., $H_2 G_2^\top = 0$). In this domain, the condition is equivalent to the requirement that each row of $H_2$ be expressible as a linear combination of the rows of $H_{\mathrm{S\text{-}ECC}}$ (i.e., $\operatorname{row}(H_2) \subseteq \operatorname{row}(H_{\mathrm{S\text{-}ECC}})$). 
If this simple condition holds, there are no additional constraints on adopting the \mytitle{} framework, which allows it to support a wide range of vendor-specific S-ECC schemes with high scalability.

\subsection{Code Construction}
\label{sec:main:code construction}
We derive an H-matrix that satisfies each layer's conditions through a two-step construction. First, we construct $H_2$ to provide SEC-DED with bounded-faults. We realize SEC-DED by assigning odd-weight columns. Instead of using a prefix region for bounded-faults, we enforce both the bounded fault and CRC8 properties by building the second half (8 columns) of each 16-column bounded region as XOR combinations of the columns in the first half (8 columns). This structure makes it easy to satisfy the CRC8 condition and also helps meet the SSC requirement of $H_{\mathrm{S\text{-}ECC}}$. 
Next, we map each bounded region of the binary $H_2$ to elements in $\text{GF}(2^{16})$ and place this symbolized $H_2$ (Fig.~\ref{fig:S-ECC1}) directly in the upper part of $H_{\mathrm{S\text{-}ECC}}$, thereby satisfying the cross-layer condition ($\operatorname{row}(H_2) \subseteq \operatorname{row}(H_{\mathrm{S\text{-}ECC}})$).

Second, we construct $H_{\mathrm{S\text{-}ECC}}$ to satisfy both SSC and DEC. Since $H_2$ is already placed in the upper part of $H_{\mathrm{S\text{-}ECC}}$, we build the lower part using a greedy search. We randomly assign $\text{GF}(2^{16})$ elements to each symbol in the lower part, then binarize $H_{\mathrm{S\text{-}ECC}}$ and check for syndrome overlaps to verify that the SSC and DEC conditions are met. If syndrome overlaps occur, we reconstruct the symbol with the largest number of overlaps and repeat this process until $H_{\mathrm{S\text{-}ECC}}$ satisfies both SSC and DEC (Fig.~\ref{fig:S-ECC2}).

\begin{figure}[t]
    \hfill
    \centering
    \includegraphics[width=\columnwidth]{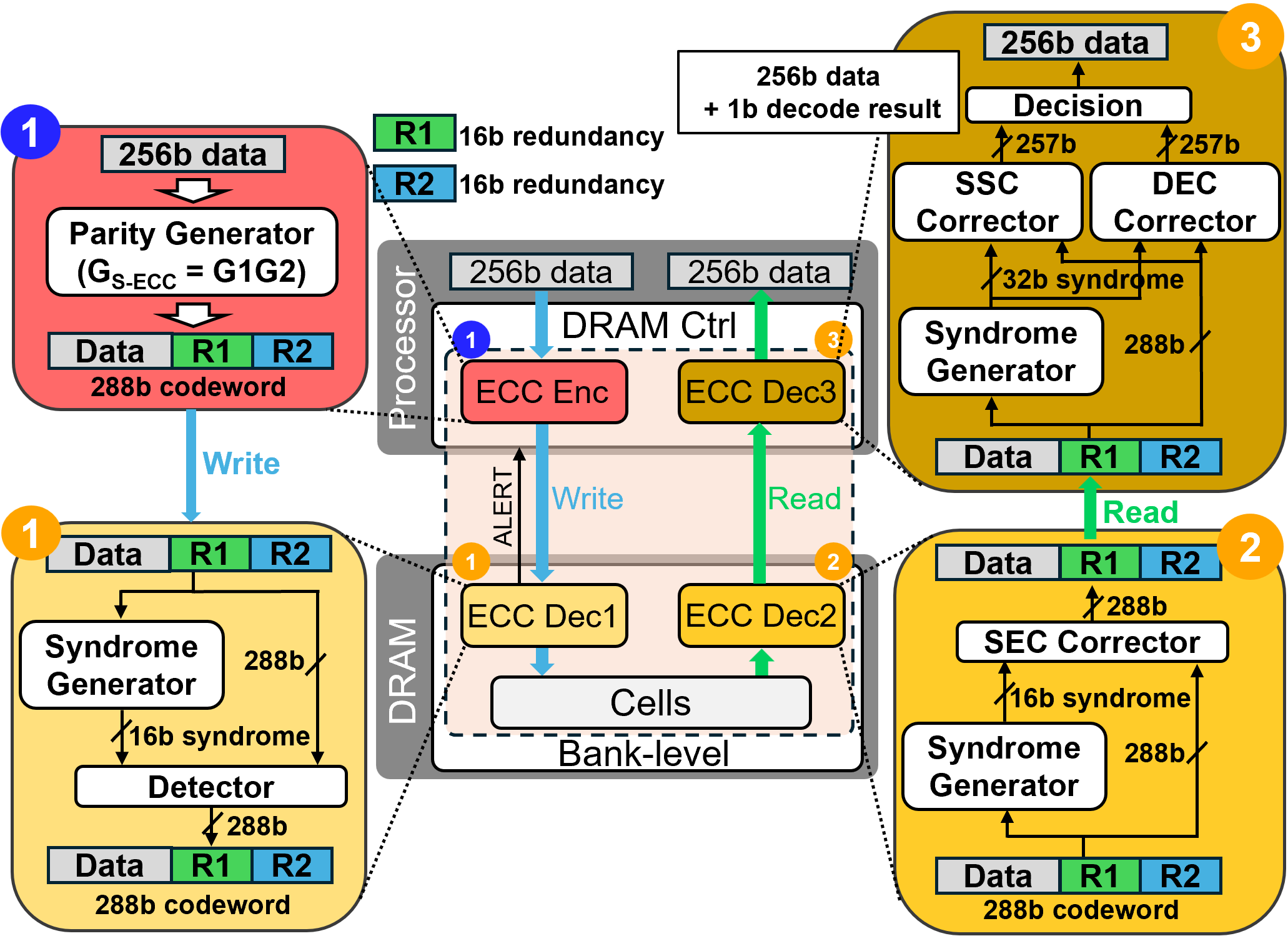}
    \caption{\ADD{The hardware implementation of \mytitle{}}}
    \label{fig:implementation}
    \vspace{-10pt}
\end{figure}

\begin{table*}[t]
\centering
\caption{A comparison of reliability against single-location error scenarios}
\label{tab:comparison_single}

\resizebox{\textwidth}{!}{
\begin{tabular}{c|c|c||c|c|c|c|c|c|c|c}
    \Xhline{3\arrayrulewidth}
    \multicolumn{3}{c||}{} 
    & \multicolumn{2}{c|}{Single-layer}
    & \multicolumn{4}{c|}{Multi-layer}
    & \multicolumn{2}{c}{Cross-layer} \\
    \cline{1-11}

    \multicolumn{3}{c||}{Redundancy (\%)} 
    & \multicolumn{1}{c|}{12.5\% (32b) }  
    & \multicolumn{1}{c|}{18.8\% (48b)}  
    & \multicolumn{2}{c|}{12.5\% (32b) }  
    & \multicolumn{2}{c|}{18.8\% (48b) }  
    & \multicolumn{1}{c|}{12.5\% (32b) }  
    & \multicolumn{1}{c} {15.6\% (40b) }  
    \\
    \Xhline{3\arrayrulewidth}
    
    \multicolumn{1}{c|}{\stackanchor{Error}{Location}} &
    \multicolumn{1}{c|}{\stackanchor{Error}{Scenario}} &
    \multicolumn{1}{c||}{\stackanchor{Decoding}{Result}} &
    \multicolumn{1}{c|}{Unity ECC} &   
    \multicolumn{1}{c|}{DUO} &          
    \makecell{\ADD{LPDDR6/}\\\ADD{SEC-DED}} &
    \makecell{\ADD{LPDDR6/}\\\ADD{CRC}} &
    \makecell{\ADD{HBM4/}\\\ADD{SEC-DED}} &
    \makecell{\ADD{HBM4/}\\\ADD{CRC}} &
    \multicolumn{1}{c|}{\mytitle{} (32b)} &
    \multicolumn{1}{c}{\mytitle{} (40b)}
    \\
    \Xhline{3\arrayrulewidth}

    \multirow{10}{*}{In bank}
    & \multirow{1}{*}{SE (\%)}
    & CE   &\multicolumn{8}{c}{100.000}  \\
    \cline{2-11}
    
    & \multirow{3}{*}{16E (\%)}
    & CE   &\multicolumn{2}{c|}{100.000} & 0.048  & 0.024  & \multicolumn{4}{c}{100.000}  \\
    && DUE &\multicolumn{2}{c|}{0.000}   & 99.563 & 99.976 & \multicolumn{4}{c}{0.000}    \\
    && SDC &\multicolumn{2}{c|}{0.000}   & 0.389  & 0.000  & \multicolumn{4}{c}{0.000}    \\
    \cline{2-11}

    & \multirow{3}{*}{32E (\%)}
    & CE   & 0.003  & 1.558  & 0.003  & 0.000  & 0.003  & 0.003  & 0.003  & 0.003 \\
    && DUE & 99.972 & 98.407 & 99.585 & 99.999 & 99.997 & 99.997 & 99.971 & 99.997 \\
    && SDC & 0.025  & 0.035  & 0.412  & 0.001  & $2\times 10^{-5}$ & $1\times 10^{-6}$ & 0.026 & $3\times 10^{-4}$\\
    \cline{2-11}

    & \multirow{3}{*}{SE+SE (\%)}
    & CE   &\multicolumn{2}{c|}{100.000} & 10.515 & 0.000 & 4.988 & 4.940 &\multicolumn{2}{c}{100.000}\\
    && DUE &\multicolumn{2}{c|}{0.000}   & 89.485 & 100.000 & 94.980 & 95.060 &\multicolumn{2}{c}{0.000}\\
    && SDC &\multicolumn{2}{c|}{0.000}   & 0.000  & 0.000   & 0.032  & 0.000  &\multicolumn{2}{c}{0.000}\\
    
    
    \Xhline{3\arrayrulewidth}
    
    \multirow{5}{*}{\stackanchor{Write}{Link}}
    & \multirow{1}{*}{SE (\%)}
    & CE    &\multicolumn{8}{c}{100.000}  \\
    \cline{2-11}

    & \multirow{1}{*}{DQE (\%)}
    & CE    &\multicolumn{8}{c}{100.000} \\
    \cline{2-11}
    
    & \multirow{3}{*}{DQSE (\%)}
    & CE    &0.000  & 0.000  & 99.998 & 99.998 & 49.994 & 49.991 &\multicolumn{2}{c}{99.999} \\
    && DUE  &99.971 & 99.945 & 0.002  & 0.002  & 50.005 & 50.008 &\multicolumn{2}{c}{0.001} \\
    && SDC  &0.029  & 0.055  & 0.000  & 0.000  & 0.001  & 0.001  &\multicolumn{2}{c}{0.000}\\
    \Xhline{3\arrayrulewidth}

    \multirow{9}{*}{\stackanchor{Out bank}{(Read peri/link)}}
    & \multirow{2}{*}{SE (\%)}
    & CE    &\multicolumn{2}{c|}{100.000} & 100.000 & 0.000 & 100.000 & 0.000 & \multicolumn{2}{c}{100.000} \\
    && DUE  &\multicolumn{2}{c|}{0.000}   & 0.000   & 100.000 & 0.000   & 100.000 & \multicolumn{2}{c}{0.000} \\
    \cline{2-11}
    
    & \multirow{2}{*}{DE (\%)}
    & CE    &\multicolumn{2}{c|}{100.000}  &\multicolumn{4}{c|}{0.000}   & \multicolumn{2}{c}{100.000}\\
    && DUE  &\multicolumn{2}{c|}{0.000}    &\multicolumn{4}{c|}{100.000} & \multicolumn{2}{c}{0.000} \\
    \cline{2-11}

    & \multirow{2}{*}{DQE (\%)}
    & CE    &\multicolumn{8}{c}{100.000} \\
    && DUE  &\multicolumn{8}{c}{0.000} \\
    \cline{2-11}
    
    & \multirow{3}{*}{DQSE (\%)}
    & CE    &0.000  & 0.000  & 99.998 & 99.998 & 49.994 & 49.991 & \ADD{99.972}  & \ADD{100.000*}\\
    && DUE  &99.971 & 99.945 & 0.002  & 0.002  & 50.005 & 50.008 & \ADD{0.000}   & \ADD{0.000}\\
    && SDC  &0.029  & 0.055  & 0.000  & 0.000  & 0.001  & 0.001  & 0.028  &$1\times 10^{-4}$\\
    \Xhline{3\arrayrulewidth}

\end{tabular}}

{\captionsetup{justification=raggedleft, singlelinecheck=false}
\caption*{\footnotesize * Rounded to 100.000\% for display; actual value is slightly lower (e.g., 99.9998\%).}}
\vspace{-15pt}
\end{table*}

\subsection{\ADD{Hardware Implementation}}
\label{sec:main:implementation}

\ADD{\mytitle{} largely reuses standard memory ECC primitives (e.g., an encoder and an SEC decoder) and updates only the $G$ and $H$ matrices (Fig.~\ref{fig:implementation}).
On the write path, the controller encoder (\circnumblue{1}) computes $R_1$ and $R_2$ in a single pass using $G_{\mathrm{S\text{-}ECC}} = G_1 \cdot G_2$. It implements this multiplication with an XOR network (e.g., 8-level XOR trees).
In DRAM, the first decoder (\circnumorange{1}) verifies writes by regenerating $R_2$ using a subset of the same XOR network.}

\ADD{On the read path, the second decoder in DRAM (\circnumorange{2}) generates a syndrome with an XOR-tree network and corrects single-bit errors based on the syndrome, which adds modest logic depth ($\approx 4$) to the combinational logic.
It then forwards the corrected 288-bit codeword (with redundancy retained) for end-to-end decoding.
The third decoder in the controller (\circnumorange{3}) applies $H_{\mathrm{S\text{-}ECC}}$ to the received 288-bit codeword to generate a 32-bit syndrome and runs SSC and DEC correctors in parallel~\cite{Kim2023Unity}. The SSC corrector uses Chien search with a modified Berlekamp--Massey procedure~\cite{Sarwate2001High}, and the DEC corrector uses a block-pair solver~\cite{Saiz-Adalid2020Reducing}.}


\ADD{Overall, the encoder and the first two decoders are on par with existing implementations. The main added complexity is the third decoder for SSC+DEC. However, it performs error \emph{detection} within a single cycle to avoid latency increases on error-free accesses, and it completes correction within a single cycle by running SSC and DEC correction in parallel. Because errors are rare, we include only the detection latency in the performance evaluation (Section~\ref{sec:eval:performance}) and report the area overhead of advanced decoding (Section~\ref{sec:eval:area}).}

\section{Evaluation}
\label{sec:eval}
We evaluate the reliability and performance impact of \mytitle{} and compare it with single-layer ECC and state-of-the-art multi-layered DRAM ECC configurations.

\subsection{Error Coverage}
\label{sec:eval:reliability}
We evaluate the reliability of \mytitle{} using Monte Carlo error-injection experiments. We inject random errors under various error scenarios, and quantify correction and detection through ECC decoding. Based on system-level analyses of DRAM faults~\cite{beigi2023systematic, wu2024removing, Jung2023Predicting, chung2025dram_fault, mbu2_sridharan2015memory}, we derive location-specific error patterns, detailed below.

We consider three error locations: (i) \emph{In bank}, (ii) \emph{Write link}, and (iii) \emph{Out bank}. 
First, \emph{In bank} covers faults internal to a DRAM bank. We consider the following error scenarios: Single Error (SE; caused by cell or BLSA), 16-bit Error (16E; CSL or SWL), 32-bit Error (32E; SWD) and a combination of two Single Errors (SE+SE). 
Second, \emph{Write link} denotes the transmit path during writes, including transmission-induced faults. We select error scenarios: SE, Data Pin Error (DQE), and Data Strobe Error (DQSE). 
Third, \emph{Out bank} spans the read path beyond the bank—device periphery and I/O—and we select error scenarios: SE, Double Error (DE; peripheral errors (e.g., TSV)) and DQE, DQSE (read-link errors).

For each error scenario, we inject errors at the specified DRAM location and flip the designated number of bits, each independently with a 50\% probability. We evaluate both single-location and multi-location cases. Each ECC scheme classifies outcomes as correctable (CE), detectable but uncorrectable (DUE), or undetectable (SDC), and we aggregate results over 10 million iterations. For link errors, we count DUEs as CEs due to retransmission and retry.

We compare \mytitle{} against a range of layered ECC configurations. For single-layer schemes, we use Unity ECC~\cite{Kim2023Unity} and DUO~\cite{gong2018DUO}, both deployed on single-device memory. For multi-layer schemes, we evaluate state-of-the-art DRAM configurations, LPDDR6~\cite{JESD209-6} and HBM4~\cite{JESD270-4}, which employ S-ECC as either SEC-DED or CRC~\cite{Hsiao1970SECDED,ryu2023IEEE}. Finally, we evaluate the cross-layer framework \mytitle{} with both 12.5\% (32b) and 15.6\% (40b) redundancy.

\subsubsection{Single-Location}Table~\ref{tab:comparison_single} summarizes error coverage and redundancy for each ECC configuration under single-location scenarios. For the \emph{In bank}, all configurations correct SE and 16E at 100\% except for 16E in LPDDR6. Because LPDDR6 partitions the 12.5\% redundancy per layer, it lacks sufficient budget to correct 16E.
\ADD{Moreover, LPDDR6/SEC-DED shows a high SDC rate for 16E and 32E due to O-ECC miscorrections.}
By contrast, \mytitle{} guarantees 100\% correction of 16E while using the same 12.5\% total redundancy budget across layers. 
For other scenarios, \mytitle{} provides strong detection capability for 32E and, with increased redundancy, can further enhance robustness. 

\begin{table*}[t]
\caption{A comparison of reliability against multi-location error scenarios}
\centering
\label{tab:comparison_multi}
\resizebox{\textwidth}{!}{
\begin{tabular}{c|c|c||c|c|c|c|c|c|c|c}
    \Xhline{3\arrayrulewidth}
    \multicolumn{3}{c||}{} 
    & \multicolumn{2}{c|}{Single-layer}
    & \multicolumn{4}{c|}{Multi-layer}
    & \multicolumn{2}{c}{Cross-layer} \\
    \cline{1-11}

    \multicolumn{3}{c||}{Redundancy (\%)} 
    & \multicolumn{1}{c|}{32b (12.5\%)}  
    & \multicolumn{1}{c|}{48b (18.8\%)}  
    & \multicolumn{2}{c|}{32b (12.5\%)}  
    & \multicolumn{2}{c|}{48b (18.8\%)}  
    & \multicolumn{1}{c|}{32b (12.5\%)}  
    & \multicolumn{1}{c} {40b (15.6\%)}  
    \\
    \Xhline{3\arrayrulewidth}
    
    \multicolumn{1}{c|}{\stackanchor{Error}{Location}} &
    \multicolumn{1}{c|}{\stackanchor{Error}{Scenario}} &
    \multicolumn{1}{c||}{\stackanchor{Decoding}{Result}} &
    \multicolumn{1}{c|}{Unity ECC} &   
    \multicolumn{1}{c|}{DUO} &          
    \makecell{\ADD{LPDDR6/}\\\ADD{SEC-DED}} &
    \makecell{\ADD{LPDDR6/}\\\ADD{CRC}} &
    \makecell{\ADD{HBM4/}\\\ADD{SEC-DED}} &
    \makecell{\ADD{HBM4/}\\\ADD{CRC}} &
    \multicolumn{1}{c|}{\mytitle{} (32b)} &
    \multicolumn{1}{c}{\mytitle{} (40b)}
    \\
    \Xhline{3\arrayrulewidth}

    \multirow{5}{*}{\makecell{In bank+\\Out bank}} 
    & \multirow{2}{*}{\makecell{SE+\\SE (\%)}} 
    & CE    &\multicolumn{2}{c|}{100.000} & 100.000 & 0.000 & 100.000 & 0.000 & \multicolumn{2}{c}{100.000} \\
    && DUE  &\multicolumn{2}{c|}{0.000}   & 0.000   & 100.000 & 0.000   & 100.000 & \multicolumn{2}{c}{0.000} \\
    \cline{2-11}
    
    & \multirow{3}{*}{\makecell{SE+\\DE (\%)}} 
    & CE  &0.947  &100.000&\multicolumn{4}{c|}{0.000} &\multicolumn{2}{c}{100.000}  \\
    & & DUE &99.019 &0.000&\multicolumn{4}{c|}{100.000} &\multicolumn{2}{c}{0.000}  \\
    & & SDC &0.034 &0.000&\multicolumn{4}{c|}{0.000} &\multicolumn{2}{c}{0.000}  \\
    \Xhline{3\arrayrulewidth}

    \multirow{6}{*}{\makecell{In bank+\\Write Link}} 
    & \multirow{3}{*}{\makecell{16E+\\DQE (\%)}} 
    & CE     &5.545  &100.000 & 0.047  & 0.025  &\multicolumn{4}{c}{100.000}  \\
    & & DUE  &94.431 &0.000   & 99.565 & 99.975 &\multicolumn{4}{c}{0.000}  \\
    & & SDC  &0.024  &0.000   & 0.388  & 0.000  &\multicolumn{4}{c}{0.000}  \\
    \cline{2-11}
    
    & \multirow{3}{*}{\makecell{32E+\\DQSE (\%)}} 
    & CE     &0.000  & 0.000  &0.002  &0.000  & 0.001 & 0.001 &0.003  &0.003  \\
    & & DUE  &99.972 & 99.950 &99.586 &99.999 & 99.999 & 99.999 &99.972  &99.997  \\
    & & SDC  &0.028  & 0.050  &0.412  &0.001  & $2\times 10^{-5}$ & $8\times 10^{-7}$ &0.025  & $3\times 10^{-4}$ \\
    \Xhline{3\arrayrulewidth}

    \multirow{6}{*}{\makecell{In bank+\\Out bank+\\Write Link}} 
    & \multirow{3}{*}{\makecell{SE+\\SE+\\SE (\%)}} 
    & CE     &1.293  &100.000 &100.000 &0.000   &100.000 &0.000  &\multicolumn{2}{c}{100.000} \\
    & & DUE  & 98.672 &0.000   &0.000   &100.000 &0.000   &100.000 &\multicolumn{2}{c}{0.000}  \\
    & & SDC  & 0.035 &0.000   &0.000   &0.000   &0.000   &0.000  &\multicolumn{2}{c}{0.000}  \\
    \cline{2-11}
    
    & \multirow{3}{*}{\makecell{SE+\\DQE+\\DQSE (\%)}} 
    & CE     & 0.000 & 0.000   & 99.998  & 99.998   & 50.010 & 49.982 & 99.998 & 99.999 \\
    & & DUE  & 97.971 & 99.950 & 0.002   &  0.002  & 49.988 & 50.017 & 0.002  & 0.001 \\
    & & SDC  & 0.028 & 0.050   & $3\times 10^{-5}$ & 0.000   & 0.001  & 0.001  & 0.000  & 0.000 \\
    \Xhline{3\arrayrulewidth}
\end{tabular}}
\vspace{-15pt}
\end{table*}


\ADD{For SE+SE, none of the multi-layer configurations provides guaranteed correction. In LPDDR6, the SEC-DED O-ECC detects the event but forwards the uncorrected data to the controller without a hint, while the downstream S-ECC (SEC-DED or CRC) lacks sufficient correction capability (except in cases where errors occur in on-die parity bits that are not transferred). Similarly, in HBM4, the SSC O-ECC corrects only when both errors fall within the same symbol; otherwise the residual pattern exceeds the S-ECC capability. This behavior aligns with prior observations on multi-layer protection~\cite{Alam2022COMET}, although the SDC rates differ because \cite{Alam2022COMET} considers SEC O-ECC (without double-error detection). Unity ECC and DUO correct SE+SE by concentrating redundancy in a strengthened single layer that supports double-bit correction. \mytitle{} also corrects SE+SE via its SSC+DEC while preserving on-die error concealment for the common case of single-bit errors.}

For the \emph{Write link}, Unity ECC and DUO can correct SE and DQE with strong S-ECC, but wider transfer errors (DQSE) are only \emph{detected}, not corrected. Because they are single-layer schemes, they cannot perform early detection, so corrupted data may be written back uncorrected, allowing subsequent faults to accumulate and increase the risk of severe reliability issues. In contrast, the multi-layer baselines employ L-ECC and thus offer high detection for SE/DQE/DQSE, enabling correction via retransmission. However, when the L-ECC is provisioned with limited redundancy, as in HBM4, DQSE detection drops to roughly half of cases. \mytitle{}, on the other hand, guarantees 100\% detection (and thus correction) for SE and DQE, and for wider DQSE, the first decoder already provides high detection; any remaining cases are caught by the stronger third decoder, yielding a robust end-to-end design.

For the \emph{Out bank}, Unity ECC and DUO can correct SE, DE, and DQE using a strong S-ECC. However, they only \emph{detect} DQSE because they do not guarantee a retry after detection. In contrast, the multi-layer configurations do not guarantee correction for peripheral DE. Moreover, when they apply S-ECC with CRC (LPDDR6-CRC and HBM4-CRC), they fail to correct even SE and instead provide detection only. This limitation stems from redundancy partitioning across layers, which prevents redundancy reuse. For example, HBM4 can allocate 32b to a strong O-ECC (SSC) and leave only 16b for S-ECC. This allocation forces weaker codes (SEC-DED or CRC) and leaves the system vulnerable to out-of-bank errors. In contrast, \mytitle{} guarantees 100\% correction for SE, DE, and DQE, \ADD{and achieves near-complete correction for DQSE with retries.}

\begin{figure*}[t]
    \centering
    \subfloat[A comparison of the IPC (normalized to HBM4, higher is better)]{
    \includegraphics[width=\textwidth]{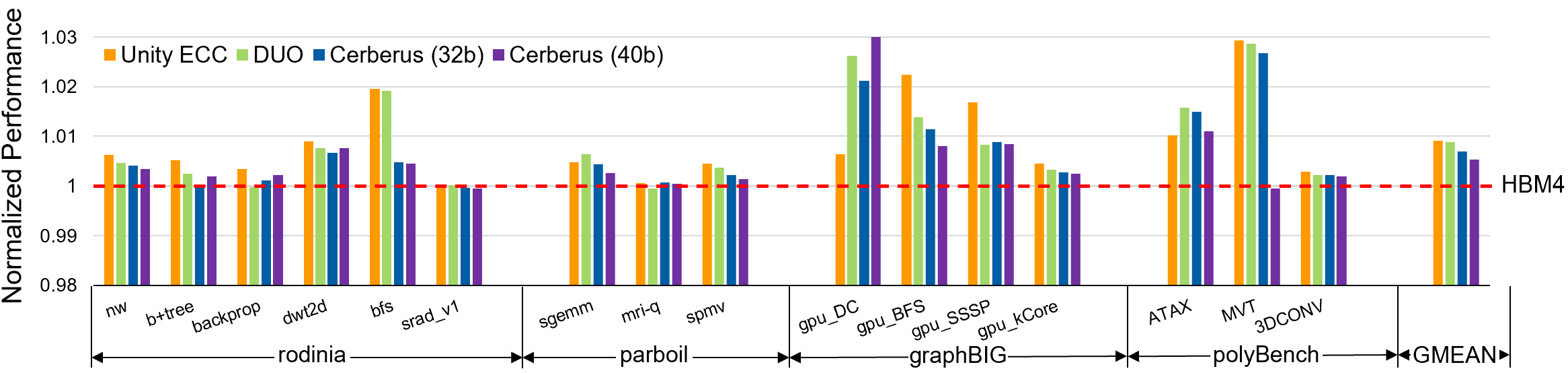}
    \label{fig:IPC}
    }

    \subfloat[\ADD{A comparison of the DRAM energy consumption (normalized to HBM4, lower is better)}]{
    \includegraphics[width=1\textwidth]{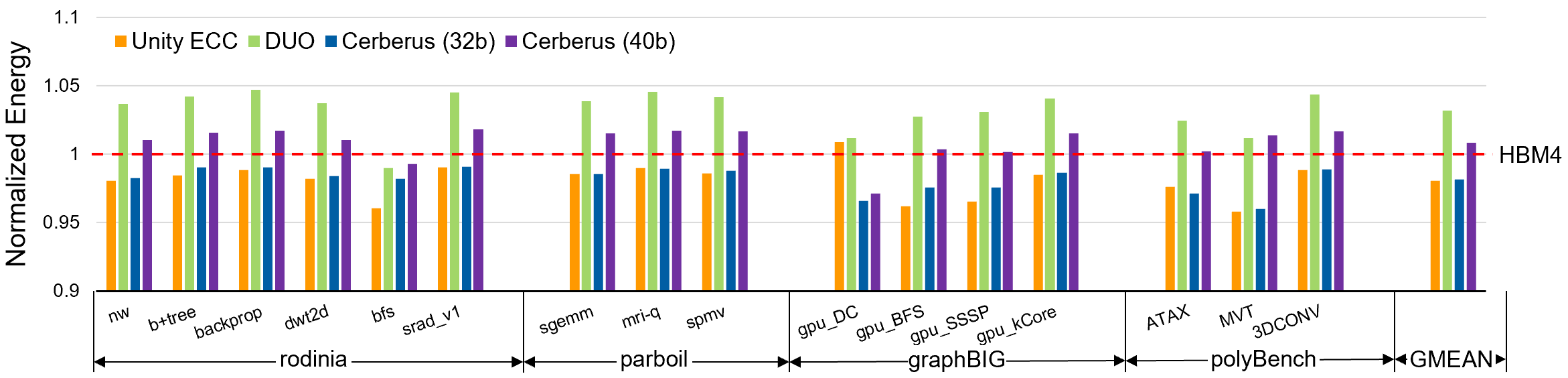}
    \label{fig:Energy}
    }
    \caption{Comparison of GPU performance and DRAM energy for \mytitle{} across the evaluated benchmarks}
    \label{fig:S-ECC}
\end{figure*}

\subsubsection{Multi-Location}Table~\ref{tab:comparison_multi} summarizes the error coverage and redundancy of each ECC configuration under multi-location scenarios. For combined \emph{In bank} and \emph{Out bank} errors, Unity ECC and DUO guarantee 100\% correction for SE+SE. However, for SE+DE, DUO still corrects while Unity ECC fails. Although Unity ECC corrects SE and DE individually in the single-location case, the overlap across two locations exposes the limitation of relying solely on S-ECC. The multi-layer approaches show the same behavior as in the single-location \emph{Out bank} case (e.g., failing to correct peripheral DE), \ADD{because although O-ECC corrects \emph{In bank} SE, \emph{Out bank} errors still remain due to the weaker code used for S-ECC.}

In contrast, \mytitle{} guarantees correction for both SE+SE and SE+DE across locations even with a small redundancy budget, \ADD{as O-ECC corrects the \emph{In bank} SE and S-ECC covers all \emph{Out bank} errors.} 
For combined \emph{In bank} and \emph{Write link}, configurations with 12.5\% redundancy (Unity ECC/LPDDR6) do not guarantee correction for 16E+DQE. However, \mytitle{} can correct it even with the same redundancy. For 32E+DQSE, \mytitle{} likewise provides significantly higher detection than other schemes with the same redundancy. 
\ADD{Finally, in the high-risk scenario where errors occur simultaneously at \emph{all} locations (\emph{In bank}, \emph{Out bank}, and \emph{Write link}), single-layer configurations reveal the limitation of relying only on S-ECC, since all overlapping errors are exposed to the system layer. DUO can correct SE+SE+SE with its stronger scheme, but it fails once larger errors are involved, and Unity ECC also cannot guarantee correction across these cases. Multi-layer configurations improve locality by letting each layer handle its corresponding errors (e.g., O-ECC handles \emph{In bank} errors and L-ECC handles link errors), but because the limited redundancy budget is divided across layers, they still fail to provide correction in all cases. In contrast, \mytitle{} retains the role of each layer while reusing redundancy across layers, enabling nearly 100\% correction across all cases with a small redundancy budget and thus offering robust end-to-end reliability.}

Rather than limiting our evaluation to in-DRAM faults, we assess reliability using error scenarios that span the entire memory system, and show that \mytitle{} maintains strong reliability under both single- and multi-location errors. In addition, \mytitle{} is a scalable framework that can accommodate higher redundancy. Although this increases overhead, it improves detection capability and yields a more robust system.

\subsection{Performance \& Energy Consumption}
\label{sec:eval:performance}
\subsubsection{GPU Performance}

We evaluate the performance impact of \ADD{\mytitle{} (32b) and \mytitle{} (40b)} on GPUs using the cycle-level simulator Accel-Sim~\cite{Khairy2020Accel-sim}, and compare it against HBM4~\cite{JESD270-4}, Unity ECC~\cite{Kim2023Unity}, and DUO~\cite{gong2018DUO}. Our system model is based on an NVIDIA V100 GPU configured with 32 HBM channels, with detailed parameters listed in Table~\ref{tab:configuration}. To capture a wide range of application behaviors, we use 16 workloads drawn from four benchmark suites: Rodinia~\cite{Che2010Rodinia}, Parboil~\cite{stratton2012parboil}, GraphBIG~\cite{Nai2015GraphBIG}, and PolyBench~\cite{Abella-Gonz2021PolyBench}.

\begin{table}[t]
  \centering
  \caption{The simulation configuration}
  \label{tab:configuration}
  \resizebox{\columnwidth}{!}{
  \begin{tabular}{c|c}
    \Xhline{3\arrayrulewidth}
    \multicolumn{1}{c|}{Components} 
    & \multicolumn{1}{c}{Configuration} \\
    \Xhline{3\arrayrulewidth}
    
    \multicolumn{1}{c|}{\# of SMs}
    & \multicolumn{1}{c}{80}\\
    \hline

    \multirow{2}{*}{SM} 
    & 1132 MHz, 4 warp schedulers/SM, \\
    & up to 32 blocks/SM, up to 48 warps/SM \\
    \hline
    
    \multicolumn{1}{c|}{L1 cache} 
    & Up to 128KiB, 4 banks, 128B line, 256-way, 384 MSHR entries\\
    \hline
    
    \multicolumn{1}{c|}{L2 cache} 
    &  4MiB, 128B lines, 16-way, 192 MSHR entries\\
    \hline

    \multirow{2}{*}{\stackanchor{Memory}{controller}} 
    & 256B channel interleaving, 64-entry scheduling queue,\\
    & FR-FCFS scheduling\\
    \hline
    
    \multirow{3}{*}{Memory} 
    & HBM4, 32 channels, 6.4Gbps\\
    & tRCD  = 30-cycle, tRRD  = 4-cycle, CL = 24-cycle, \\
    & tCCDS = 2-cycle,  tCCDL = 4-cycle, WL = 14-cycle\\
    \hline
    
    \Xhline{3\arrayrulewidth}
  \end{tabular}}
\end{table}

The ECC decoder affects the time from a read command to the output of the first corresponding data beat (tCL), while the encoder impacts the write latency (tWL)~\cite{Cha2017}. We evaluate the performance by adjusting these two timing parameters for each ECC configuration. For HBM4, we estimate the encoder and decoder latency overheads of 16-bit SSC-based O-ECC as 2ns and 5ns, respectively~\cite{Cha2017}. In contrast, the latency overheads of both S-ECC (CRC) and L-ECC (parity) are less than 1ns for both encoding and decoding. 
For Unity ECC and DUO, we derive their timing parameters by first removing the O-ECC latency overhead from the HBM4 baseline and then adding each scheme’s synthesized latency. Logic synthesis in a UMC 28nm process yields latency overheads of 1.46ns for Unity ECC and 1.92ns for DUO. 
\ADD{For both \mytitle{} (32b) and \mytitle{} (40b), we remove the O-ECC encoder latency and reduce the decoder latency from 5ns to 2ns under the modified O-ECC scheme~\cite{Cha2017}, since the two configurations have nearly identical O-ECC latency. We also model S-ECC latency from logic synthesis, adding 0.85ns for \mytitle{} (32b) and 0.89ns for \mytitle{} (40b).} Finally, we convert all latency values into clock cycles at 1.6~GHz (6.4~Gbps after QDR in HBM4).

Fig.~\ref{fig:IPC} reports the instructions per cycle (IPC) for HBM4, Unity ECC, DUO, \ADD{\mytitle{} (32b) and \mytitle{} (40b)}, normalized to HBM4. \ADD{Across benchmarks, \mytitle{} (32b) improves IPC by 0.2\%, 0.2\%, 1.1\%, and 1.4\% (0.7\% geomean), while \mytitle{} (40b) improves IPC by 0.3\%, 0.1\%, 1.2\%, and 0.4\% (0.5\% geomean).} Unity ECC and DUO achieve similar IPC gains, but they provide lower reliability than \mytitle{}. In contrast, \mytitle{} delivers higher reliability than HBM4 while using less redundancy, yet still improves IPC. It also attains performance comparable to Unity ECC (0.9\% geomean). These benefits stem from the EODM organization of \mytitle{}, which eliminates repeated encoding stages through efficient reuse of redundancy across layers.

\begin{table}[t]
\caption{\ADD{Estimated DRAM operating currents (per pseudo-channel)}}
\label{tab:dram_currents}
\resizebox{\columnwidth}{!}{
\begin{tabular}{c|c|c|c|c|c}
    \Xhline{3\arrayrulewidth}
    \makecell{DRAM current} & \makecell{Unity\\ ECC} & DUO & HBM4 & \makecell{Cerberus\\ (32b)} & \makecell{Cerberus\\ (40b)} \\
    \Xhline{3\arrayrulewidth}
    
    \makecell{IDD0 (mA)\\(ACT-PRE)} &
    \makecell{ 47.42\\ (99.7\%)} &
    \makecell{ 47.56\\ (100\%)} &
    \makecell{ 47.56\\ (100\%)} &
    \makecell{ 47.42\\ (99.7\%)} &
    \makecell{ 47.49\\ (99.9\%)} \\
    \hline

    \makecell{IDD2N (mA)\\(Precharge standby)} 
    & \multicolumn{5}{c}{\makecell{ 44.88\\ (100\%)}} \\
    \hline
    
    \makecell{IDD3N (mA)\\(Active standby)} &
    \makecell{ 47.69\\ (99.7\%)} &
    \makecell{ 47.84\\ (100\%)} &
    \makecell{ 47.84\\ (100\%)} &
    \makecell{ 47.69\\ (99.7\%)} &
    \makecell{ 47.77\\ (99.8\%)} \\
    \hline
    
    \makecell{IDD4R (mA)\\(Read)} &
    \makecell{521.28 \\ (99.2\%)} &
    \makecell{547.75 \\ (104.2\%)} &
    \makecell{525.72 \\ (100\%)} &
    \makecell{521.28 \\ (99.2\%)} &
    \makecell{534.52 \\ (101.7\%)} \\
    \hline
    
    \makecell{IDD4W (mA)\\(Write)} &
    \makecell{365.27 \\ (99.2\%)} &
    \makecell{383.07 \\ (104\%)} &
    \makecell{368.25 \\ (100\%)} &
    \makecell{365.29 \\ (99.2\%)} &
    \makecell{374.17 \\ (101.6\%)} \\
    
    \Xhline{3\arrayrulewidth}
\end{tabular}
}
\end{table}

\subsubsection{DRAM Energy Consumption}

\ADD{\mytitle{} uses different storage and transfer bit widths than HBM4. To estimate the resulting DRAM power/energy, we use HBM2E operating currents from a datasheet~\cite{SamsungHBM2E2021}.}

\ADD{We assume the precharge standby current (IDD2N) is independent of bit width, while the incremental activation current (IDD0$-$IDD2N) and active-standby current (IDD3N$-$IDD2N) scale with the stored bit count (e.g., $(256{+}32{+}16)$ bits in HBM4 vs.\ $(256{+}32)$ bits in \mytitle{} (32b)).
For read/write activity, we partition the incremental currents (IDD4R$-$IDD2N and IDD4W$-$IDD2N) between bank-group-internal transfer (cells$\rightarrow$O-ECC) and bank-group-external transfer (O-ECC$\rightarrow$processor) and apply a 61\%:39\% split from prior HBM2 analysis~\cite{o2017fine}. We then scale each component by the corresponding transfer width: HBM4 transfers $(256{+}32{+}16)$ bits within a bank group and $(256{+}16)$ bits outside, whereas \mytitle{} (32b) transfers $(256{+}32)$ bits within a bank group and $(256{+}32)$ bits outside. 
Finally, we compute overall DRAM energy consumption using the Micron DDR4 power calculator~\cite{micron2017tn} with these current values. Table~\ref{tab:dram_currents} summarizes the DRAM operating currents used in our evaluation.}

\ADD{Fig.~\ref{fig:Energy} shows the results. \mytitle{} (32b) reduces energy by 1.84\% on average compared to HBM4. This is primarily because intra-die transfers from cells to bank peripherals often consume more energy than off-chip transfer~\cite{o2017fine, chatterjee2017architecting}, and \mytitle{} reduces these bank-group--internal transfers. With higher redundancy, \mytitle{} (40b) consumes 0.86\% more energy than HBM4 on average to provide stronger protection.}

\subsection{\ADD{Hardware Overheads}}
\label{sec:eval:area}

\ADD{To estimate the hardware costs, we implement SystemVerilog models for the encoder and decoders. We synthesize these models with Synopsys Design Compiler using a UMC 28nm standard library. We then normalize the resulting area to NAND2 equivalents (the number of NAND2 gates that occupy the same area) to present process-independent results.}


\ADD{Table~\ref{tab:area_overhead} summarizes the area cost of \mytitle{}. Decoder~1 (L-ECC) and Decoder~2 (O-ECC) reside inside the DRAM device. Together, they require only 14,693 NAND2 equivalents in \mytitle{} (32b) and 15,662 NAND2 equivalents in \mytitle{} (40b), which correspond to 0.0074~mm$^2$ and 0.0079~mm$^2$, respectively. This overhead is negligible compared to an HBM stack footprint (e.g., 121~mm$^2$ for HBM3~\cite{Duplex2024Yun}). Decoder~3 (S-ECC) dominates the overall overhead, mainly due to the DEC corrector. Even so, the processor-side overheads (encoder+S-ECC) total 127,583 NAND2 equivalents for \mytitle{} (32b) and 167,954 NAND2 equivalents for \mytitle{} (40b). Relative to modern GPUs with billions of transistors, this is a tiny fraction of the transistor count (e.g., $2.5\times 10^{-6}$ to $3.2\times 10^{-6}$ of a 208B-transistor Blackwell~\cite{Micro2026Aaron}).}

\begin{table}
 \caption{\ADD{Area overheads (in NAND2 equivalents)}}
 \label{tab:area_overhead}
 \small
 \resizebox{\columnwidth}{!}{
 \begin{tabular}{c|c|c}
    \Xhline{3\arrayrulewidth}
    & \makecell{Cerberus (32b)} & \makecell{Cerberus (40b)} \\
    \Xhline{3\arrayrulewidth}

    Encoder 
    &  1632.79 $\mu\mathrm{m}^2$ (3240)
    &  2015.49 $\mu\mathrm{m}^2$ (3999)
    \\
    \hline
    \makecell{Decoder 1\\(L-ECC)}
    &  1205.40 $\mu\mathrm{m}^2$ (2392)
    &  1398.43 $\mu\mathrm{m}^2$ (2775)
    \\
    \hline
    \makecell{Decoder 2\\(O-ECC)}   
    &  6199.87 $\mu\mathrm{m}^2$ (12301)
    &  6495.15 $\mu\mathrm{m}^2$ (12887)
    \\
    \hline
    \makecell{Decoder 3\\(S-ECC)}   
    &  62669.04 $\mu\mathrm{m}^2$ (124343)
    &  82633.15 $\mu\mathrm{m}^2$ (163955)
    \\

    \Xhline{3\arrayrulewidth}
\end{tabular}
}
\end{table}

\vspace{10pt}

\section{Conclusion}
\label{sec:conclusion}

This paper presents \mytitle{}, a cross-layer ECC co-design that addresses key challenges of multi-layer ECC: inefficient use of redundancy, overlapping protection coverage, and destructive cross-layer interference due to miscorrections. Implemented on HBM4 with a cross-layer ECC design, \mytitle{} reduces redundancy by 33.3\% while still providing higher reliability through efficient redundancy reuse. Moreover, its Encode-Once, Decode-Many (EODM) architecture eliminates unnecessary encoding stages, improving performance and delivering seamless coverage without protection gaps. Overall, \mytitle{} provides a promising framework for achieving high reliability in future HBM- and LPDDR-based systems.

\bibliographystyle{IEEEtran}
\bibliography{References}

\end{document}